\documentclass[12pt,preprint]{aastex}

\usepackage{graphicx}
\usepackage{epstopdf}

\newcommand{\dn}{~D$_n$(4000)~}
\newcommand{\hda}{~H$\delta_A$~}
\newcommand{\hess}{Hess function}
\newcommand{\area}{524}

\newcommand{\fuvmag}{\ifmmode{FUV}\else{\it FUV~}}
\newcommand{\nuvmag}{\ifmmode{NUV}\else{\it NUV~}}

\slugcomment{Submitted for publication in the Special GALEX Ap.J.Suppl. Issue}

\shorttitle{UV Contraints on Evolution from Blue to Red Sequence}
\shortauthors{Martin et al.}

\begin{document}

\title{The Ultraviolet-Optical Hess Diagram III: Constraints on Evolution from the Blue to Red Sequence}

\author{
D. Christopher Martin\altaffilmark{1},
Ted K. Wyder\altaffilmark{1},
David Schiminovich\altaffilmark{10},
Tom A. Barlow\altaffilmark{1},
Karl Forster\altaffilmark{1},
Peter G. Friedman\altaffilmark{1},
Patrick Morrissey\altaffilmark{1},
Susan G. Neff\altaffilmark{8},
Mark Seibert\altaffilmark{1},
Todd Small\altaffilmark{1},
Barry Y. Welsh\altaffilmark{6}, 
Luciana Bianchi\altaffilmark{2},
Jose Donas\altaffilmark{4},
Timothy M. Heckman\altaffilmark{5},
Young-Wook Lee\altaffilmark{3},
Barry F. Madore\altaffilmark{7},
Bruno Milliard\altaffilmark{4},
R. Michael Rich\altaffilmark{9},
Alex S. Szalay\altaffilmark{5},
Sukyoung K. Yi\altaffilmark{3}
}

\altaffiltext{1}{California Institute of Technology, MC 405-47, 1200 East
California Boulevard, Pasadena, CA 91125}

\altaffiltext{2}{Center for Astrophysical Sciences, The Johns Hopkins
University, 3400 N. Charles St., Baltimore, MD 21218}

\altaffiltext{3}{Center for Space Astrophysics, Yonsei University, Seoul
120-749, Korea}

\altaffiltext{4}{Laboratoire d'Astrophysique de Marseille, BP 8, Traverse
du Siphon, 13376 Marseille Cedex 12, France}

\altaffiltext{5}{Department of Physics and Astronomy, The Johns Hopkins
University, Homewood Campus, Baltimore, MD 21218}

\altaffiltext{6}{Space Sciences Laboratory, University of California at
Berkeley, 601 Campbell Hall, Berkeley, CA 94720}

\altaffiltext{7}{Observatories of the Carnegie Institution of Washington,
813 Santa Barbara St., Pasadena, CA 91101}

\altaffiltext{8}{Laboratory for Astronomy and Solar Physics, NASA Goddard
Space Flight Center, Greenbelt, MD 20771}

\altaffiltext{9}{Department of Physics and Astronomy, University of
California, Los Angeles, CA 90095}

\altaffiltext{10}{Department of Astronomy, Columbia University,
New York, New York 10027}


\begin{abstract}

We introduce a new quantity, the mass flux density of galaxies evolving
from the blue sequence to the red sequence. We propose a simple technique
for constraining this mass flux using the volume corrected number density
in the extinction-corrected UV-optical color magnitude distribution, the stellar age indexes \hda and \dn,
and a simple prescription for spectral evolution using a quenched star formation history.
We exploit the excellent separation of red and blue sequences in the NUV-r band \hess. 
The final value we measure, $\dot{\rho_T}=0.033$ M$_\odot$ yr$^{-1}$ Mpc$^{-3}$, is strictly
speaking an upper limit due to the possible contributions of bursting,
composite, and extincted galaxies. However, it compares favorably with
estimates of the average mass flux that we make based on the red luminosity function evolution
derived from the DEEPII and COMBO-17 surveys \citep{bell04b,faber05},
$\dot{\rho}_R=+0.034$ M$_\odot$ yr$^{-1}$ Mpc$^{-3}$.
We find that the blue sequence mass has remained roughly constant
since z=1 ($\dot{\rho_B}\simeq 0.01$ M$_\odot$ yr$^{-1}$ Mpc$^{-3}$,
but the average on-going star formation of 
$\dot{\rho}_{SF}\simeq0.037$ M$_\odot$ yr$^{-1}$ Mpc$^{-3}$ over $0<z<1$ is balanced
by mass flux off the blue sequence.
We explore the nature of the galaxies in the transition zone with particular attention
to the frequency and impact of AGNs. The AGN fraction peaks in the transition zone.
We find circumstantial, albeit weak evidence that the quench rates are higher in higher luminosity AGNs.

\end{abstract}



\section{Introduction}

There has been growing interest in the nature of the observed
color bimodality in the distribution of galaxies \citep{balogh04,baldry04}, which is
echoed in other galaxy properties \citep{kauffmann03}. The color bimodality is
revealed in a variety of color-magnitude plots, and is particularly dramatic
in the UV-optical color magnitude diagram \citep{wyder06}. The red and blue galaxy concentrations
have been denoted the red sequence and the blue ``cloud'', although we
elect to call both concentrations sequences. Deep galaxy surveys are now
probing the evolution of the red and blue sequences.  Recent
work using the COMBO-17 \citep{bell04b} and DEEP2 \citep{faber05} surveys
provides evidence that the red sequence has grown in mass
by a factor of three since z$\sim$1. It is natural to ask what processes
have led to this growth, and in particular whether the red sequence
has grown via gas rich mergers, gas-less (dry) mergers, or simple gas
exhaustion, and whether AGN feedback has played a role in accelerating
this evolution \citep{springel05,dimatteo05}. 

We would like very much to identify galaxies which may be
in the process of evolving from the blue to the red sequence. 
In this paper, we exploit the pronounced demarcation between blue and red
sequences afforded by the Near UV-optical color-magnitude diagram 
\cite{wyder06}. We make the natural assumption that galaxies
residing in the region between the blue and red sequences are 
making a transition from the former to the latter. We develop
a metric (albeit model-dependent) for measuring the speed galaxies
are evolving through this transition zone using spectroscopic indices. 
We use this to estimate the total mass flux of galaxies from the blue to the red sequence
during the recent several gigayears. Remarkably, the result we obtain
is consistent with estimates based on the evolution of both the red 
and the blue sequences taken separately.

\section{Methodology}

We seek to determine the total rate that galaxies residing on the blue sequence
evolve to the red sequence. Our approach is to make a series of simplifying
assumptions, which we examine in section \S\ref{sec_discussion}.

\begin{enumerate}

\item Galaxies evolving from the blue to red sequence pass relatively smoothly
through intermediate colors. Galaxies do not ``jump'', for example by hiding due to
strong extinction only to be unvealed as a fully formed red sequence galaxy. In other
words we assume that galaxies transition on timescales no shorter than the
precision of the tracers we use below to determine those timescales.

\item Galaxies can be described on-average in a given intermediate
color-magnitude bin by a single parameterizable star formation history. In other words,
we ignore composite stellar populations.

\item We can ignore for the moment the detailed accounting of 
stellar mass growth due to merging. Rather, our goal is to determine
the mass flux off the blue sequence, as a function of stellar mass, and 
then to total the mass flux leaving the blue sequence.

\item The contribution of galaxies bluing (e.g., due to starbursts) from red to blue sequences
through intermediate color bins can be ignored. 

\end{enumerate}

With these assumptions, our methodology is quite simple. We define
$\dot{\phi}_{B}(M_r)$ as the number flux per unit volume at a given absolute
magnitude $M_r$ off the blue-sequence (unless explicitely stated the
sign of all mass fluxes are defined to be positive). Given that
we choose a narrow range of NUV-r color, there is a close
correspondance between $M_r$ and stellar mass.
Then the fundamental governing equation is

\begin{equation}
{\phi}(M_r, NUV-r) = \dot{\phi}_{B}(M_r) <\tau(M_r, NUV-r; \xi)>
\end{equation}

where ${\phi}(M_r, NUV-r)$ is the volume-corrected NUV-r, M$_r$ color-magnitude function (\hess),
and $\tau(M_r, NUV-r; \xi)$ is the lifetime of the galaxy in this particular color-magnitude bin, for
a star formation history $\xi$.

For most physically reasonable star formation histories, the lifetime in the bin
is determined by the rate of change of the color, such that:

\begin{equation}
{\phi}(M_r, NUV-r) \simeq  { \dot{\phi}_{B}(M_r) \over {dy/dt(y;\xi)}}
\end{equation}

where $y=NUV-r$. $dy/dt$ is positive due to assuption 4. The total mass flux is then

\begin{equation}\label{eqn_sum}
\dot{\rho}_{B\rightarrow R} \le {\sum_{M_r}}~ M(M_r)~ {\phi}(M_r, NUV-r)~ <{dy \over dt}[y;\xi]>
\end{equation}

where $M(M_r)$ is the stellar mass corresponding
to $M_r$. This equation is an equality only if all galaxies present in the intermediate
color bin are moving from the blue to red sequence. Otherwise
it represents an upper limit.

We have determined ${\phi}(M_r, NUV-r)$ in \cite{wyder06}. The key remaining
issue is to determine $dy/dt$. Our approach will be to posit 
some simple quenched star formation histories and calculate
stellar age indexes such as $D_n(4000)$ for comparison with the
observed galaxy indexes. These age measures will help constrain
the family of possible quenching histories at the transitional colors.

\section{Data}

\subsection{The Hess Function}

We generate the volume-corrected NUV-r vs. M$_r$ color-magnitude diagram
as discussed in \cite{wyder06}. We briefly recapitulate the
basic method. Our sample is NUV selected in the GALEX Medium
Imaging Survey (MIS). The MIS/SDSS DR4 co-sample occupies
\area sq. deg. of the north galactic polar cap and the southern
equatorial strip. Our sample is cut as follows:
1) NUV detection, nuv\_weight $>$ 800;
2) SDSS main galaxy sample, $z_{conf}>0.67$ and specclass=2;
3) $14.5<r_0<17.6$, $16<NUV<23.0$;
4) nuv\_artifact$<$2;
5) field radius less than 0.55 degrees;
6) $0.02<z<0.22$.
We generate the volume correction using a simple $V_{max}$ approach, with
joint selection in NUV and r magnitude. The NUV and r-band magnitudes
are k-corrected to $z=0.1$ using the \cite{blanton03} code. We
have made no evolutionary correction.

As shown by \cite{wyder06} the \hess~ displays a well-defined blue sequence
and red sequence. The minimum $NUV-r$ color between the two sequences varies slightly with
$M_r$, but a good overall choice is $NUV-r\simeq 4$. More precisely, unless otherwise
stated we use the bin with $4.0 < NUV-r < 4.5$, or $NUV-r\simeq 4.25$. As we discuss in section 6.1.5,
the results do not depend significantly on this choice.

\subsection{Extinction Corrections\label{sec_extinction_correction}}

The NUV-r color of galaxies, and to a lesser extent $M_r$, is affected by dust extinction
as well as stellar age (e.g.,\cite{johnson06,wyder06}. We therefore 
construct an extinction-corrected \hess~ in order
to determine the number density of transition galaxies. 
We do not use the method of \cite{johnson06} because we use
the combination of \dn and NUV-r later as
a transition rate metric. \cite{kauffmann03} measured dust extinction in the z-band (A$_z$)
using the continuum fluxes. \cite{brinchmann04} measure extinction using
the Balmer decrement. Using the derived A$_z$ is not ideal because they were 
derived by comparing \dn, \hda and the continuum fluxes to
a large suite of possible star formation histories.
As we discuss in the next section, we also use these indices. 
We have used both methods to correct the \hess~
and find that they yield very similar results (cf. \S\ref{sec_discussion}).
We discuss there why using the derived A$_z$ is a reasonable approach. 
Our baseline correction uses the \cite{kauffmann03} A$_z$. Using a
\cite{calzetti94} extinction law, we find that $A_r=1.6 A_z$ and
$A_{NUV}=3.8 A_z$. 

\begin{figure}
\plottwo{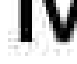}{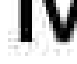}
\caption{LEFT: Hess diagram with colors indicating mean b/a (minor to major axis ratio)
for galaxies in each color-magnitude bin. 
In this and subsequent
Hess plots the fractional dispersion in the quantity
displayed as a color is given in each bin. The fractional dispersion is the
actual dispersion normalized by the range given in the color bar, i.e.,
0.4 is 40\% fractional dispersion over a range in b/a of 0.28, or 0.11 This diagram shows that extinction
in edge-on galaxies produces a significant bias in the diagram. RIGHT:
Extinction corrected Hess diagram (using continuum derived A$_z$) with
mean b/a in each bin. Bias observed in left hand plot is now virtually eliminated.
\label{fig_bova}}
\end{figure}

We show evidence that the extinction correction is working reasonably in Figure \ref{fig_bova}.
This shows the \hess~ overlaid with the mean minor to major axis ratio for each bin.
Without extinction correction (left panel), there is a clear systematic trend of lower inclination
galaxies falling in redder NUV-r bins, since lower inclination galaxies
will exhibit higher dust extinctions. After the correction (right panel), this trend is almost
completely eliminated. 

As we see below, the extinction correction has two effects. It reduces the
volume density of galaxies in the transition zone. Moreover it significantly
reduces a population of transition color galaxies showing younger than expected
spectroscopic ages.

\subsection{Star Formation History Indicators}

\begin{figure}
\plottwo{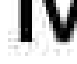}{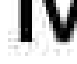}
\caption{Contours show the volume corrected NUV-r, M$_r$ color-magnitude diagram (\hess). 
Contours are spaced logarithmically in 10 even steps from $\phi=10^{-5}-10^{-3}$ (0.2 dex per step).
Colors indicate the mean value of \hda in each color-magnitude bin. The numeric value in each bin
is the fractional standard deviation of \hda in each bin, or the standard deviation divided by
the typical range of \hda. LEFT: no extinction correction; RIGHT: extinction correction. \label{fig_hess_hda}}
\end{figure}

\begin{figure}
\plottwo{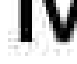}{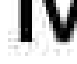}
\caption{Contours show the volume corrected NUV-r, M$_r$ color-magnitude diagram (\hess). 
Contours are spaced logarithmically in 10 even steps from $\phi=10^{-5}-10^{-3}$ (0.2 dex per step).Colors
indicate the mean value of \dn in each color-magnitude bin. The numeric value in each bin
is the fractional standard deviation of \dn in each bin, or the standard deviation divided by
the typical range of \dn. LEFT: no extinction correction; RIGHT: extinction correction. \label{fig_hess_d4000}}
\end{figure}

We use \dn and \hda as calculated and employed for the SDSS
spectroscopic sample by \cite{kauffmann03} and available as the MPA/JHU DR4 Value-Added
Catalog. \hda is corrected for nebular emission. We calculate the mean
and standard deviation of these quantities (when measured) 
in each color-magnitude bin of the uncorrected \hess.
In Figure \ref{fig_hess_d4000} and \ref{fig_hess_hda} we show the mean values of \dn and \hda 
in each color-magnitude bin (shown by the bin's color), with the density contours of the \hess~ superimposed. 
We show this for both the observed colors and the extinction-corrected colors.
Average values of the two indices transition smoothly from younger to older
characteristic ages moving from the blue to the red sequence at a given $M_r$.
For a given $NUV-r$, ages decrease slowly with decrease r-band luminosity.
The relative dispersion of the index within each bin is also shown. Typically,
the dispersions are $\sim$0.3 for \dn (of the full range of about 1-2.5) and $\sim 2$ for \hda (of the full range of -2.5 to 7 in the mean).

\begin{figure}
\plottwo{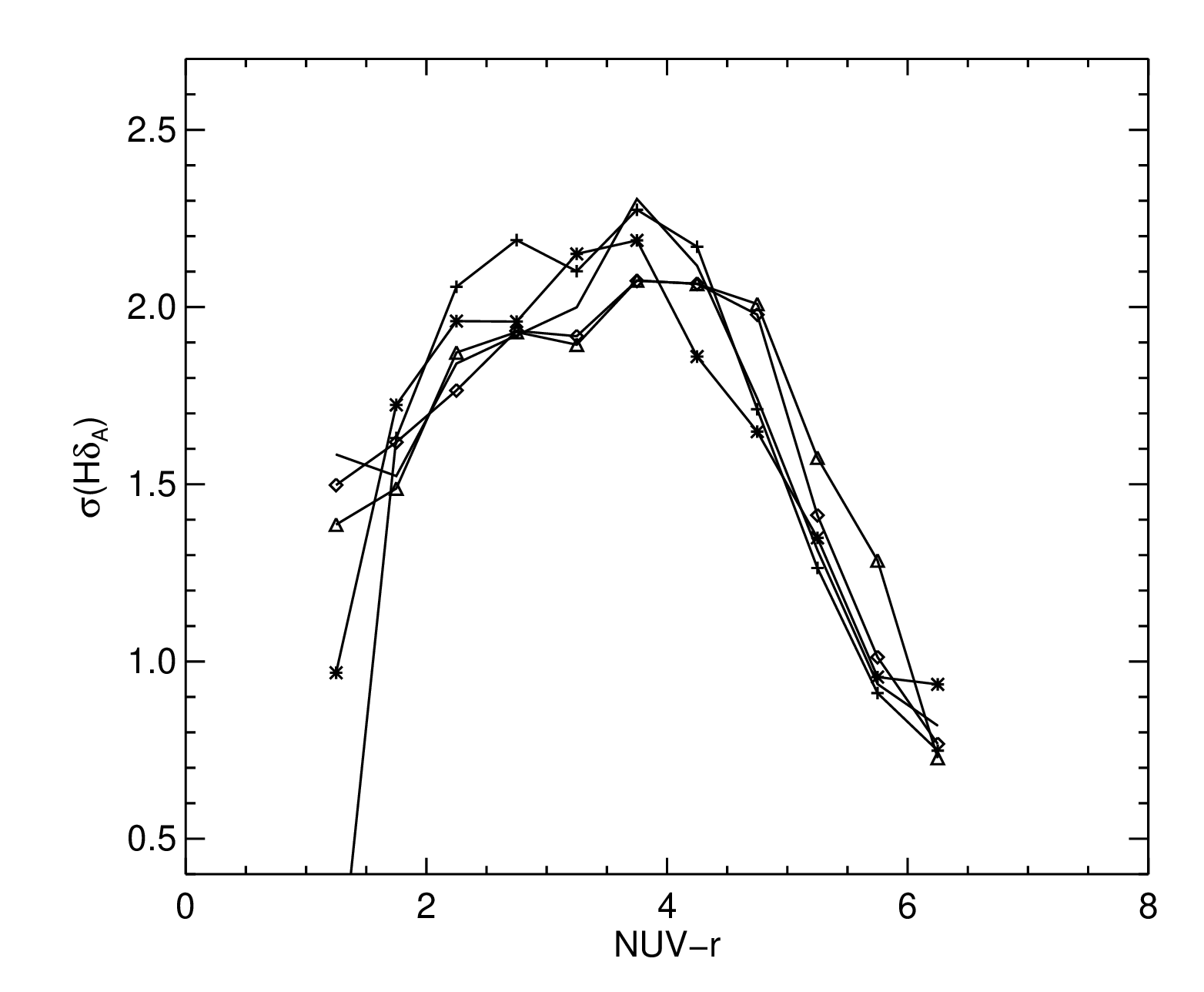}{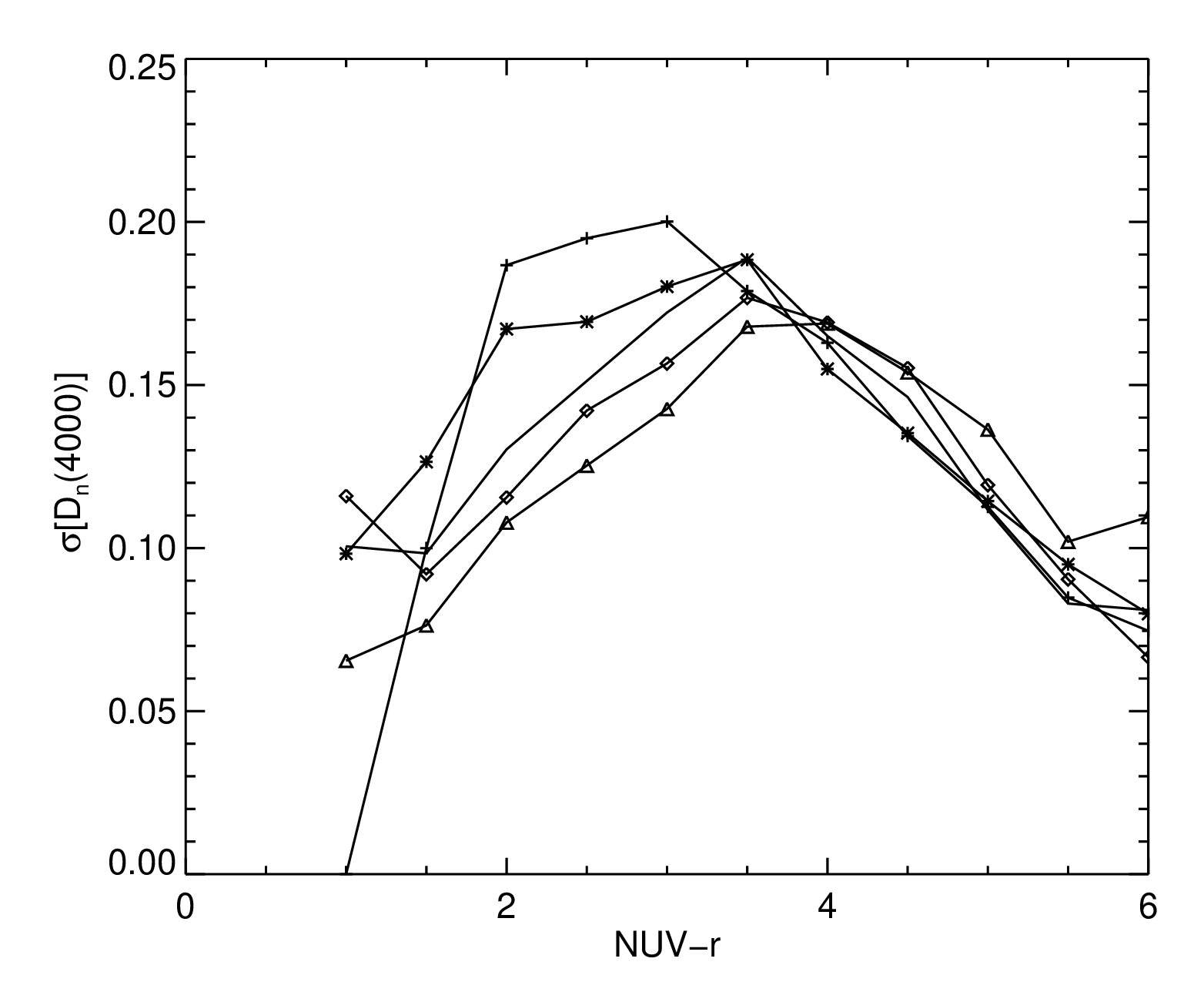}
\caption{Plots show variation of standard deviation from color-magnitude bin means
of \hda (LEFT) and \dn (RIGHT), vs. NUV-r. Symbols give variation with M$_r$,
with $M_r=-22.25,-21.75,-21.25,-20.75,-20.25$ given by symbol: plus, cross, dot, diamond, and triangle.
\label{fig_sigma}}
\end{figure}

\begin{figure}
\plottwo{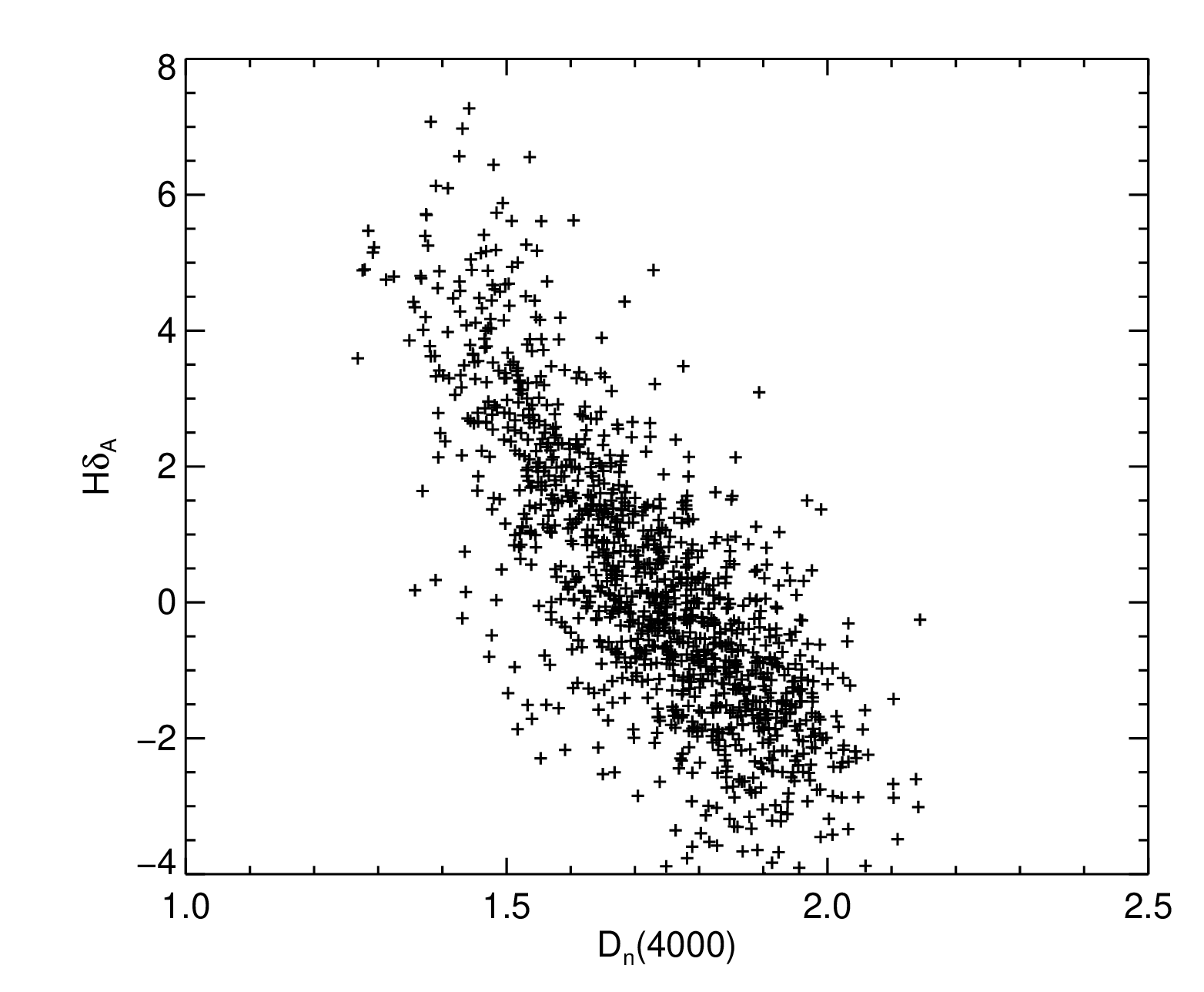}{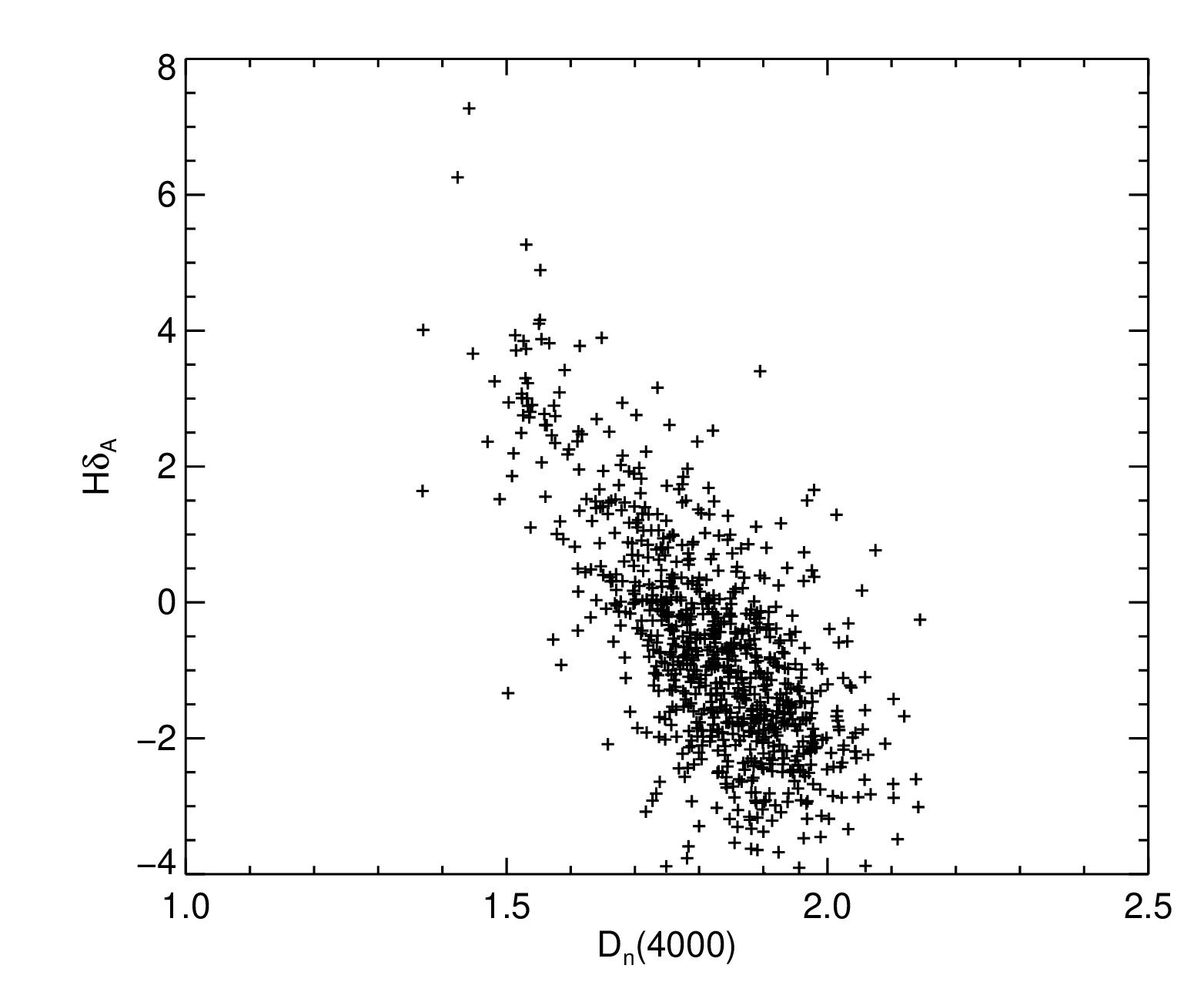}
\caption{Covariance
of \hda  and \dn in a single color-magnitude bin
(M$_r$=-21, NUV-r=4.25). LEFT: no extinction correction,
RIGHT: A$_z$ extinction correction.
\label{fig_hda_vs_d4000_dat}}
\end{figure}

The dispersion in both quantities is higher in the transition zone.
This is illustrated in Figure \ref{fig_sigma}. In Figure \ref{fig_hda_vs_d4000_dat}
we show that the variance in \hda and \dn is correlated as would be expected
(and is not due to measurement error).
We show these for both the uncorrected and extinction-corrected
\hess. The dispersion is somewhat lower in the extinction-corrected case, as some of the
lower stellar age galaxies have been shifted to bluer NUV-r bins. 
The dispersion suggests that we find a range of star formation histories in a
given NUV-r, M$_r$ color-magnitude bin, even after correction for extinction. 
We explore this issue in the next section.

\section{Quenched Star Formation Models}

We seek to explore the rate at which galaxies are transitioning from the blue to the red sequence.
In order to complete this analysis, we need to determine the color derivative $dy/dt$ in the transition region.
We also hope to gather additional evidence that our star formation history models
are a plausible description of the transition galaxies.

Our approach is to use a very simple, parametric model.
We assume that star formation on the blue sequence proceeds at a constant rate, and is then
quenched at a time $t_q$. We allow a single free parameter, the quenching rate $\gamma$. Thus the star formation history we assume
is simply:

\begin{eqnarray}\label{eqn_qsfh}
\dot{M}_* & = & \dot{M}_*(0) (t<t_q) \\
\dot{M}_* & = & \dot{M}_*(0)\left( 1-e^{-\gamma (t-t_q)} \right) (t\ge t_q)
\end{eqnarray}

We use the \cite{bruzual03} stellar synthesis models to calculate the evolution of the spectral energy
distribution. We assume solar metallicity and
Salpeter initial mass function. The metallicity assumption is reasonable
because, although there is a mass-metallicity variation, the bulk of the mass
flux comes from masses $11< \log{M}<10$, where metallicities are near solar.
The small variations around this produce even smaller variations in the model tracks. 
We also use unextinguished models. 
We assume that $t_q=10$ Gyr, which is long after all the indexes and colors we consider have
reached stable values. We examine below the impact of this assumption.

\begin{figure}
\plotone{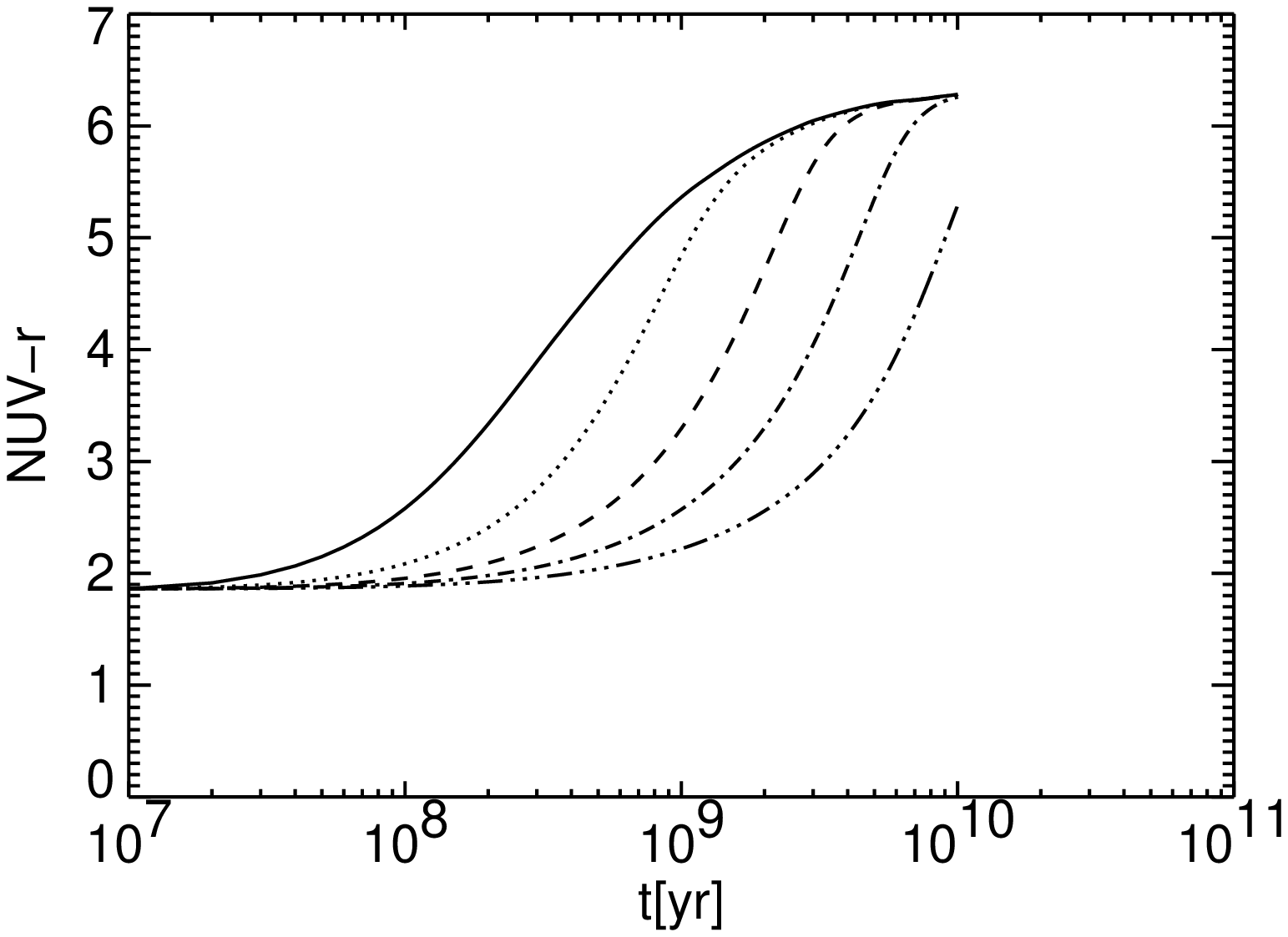}
\caption{Quenched star formation history, NUV-r vs. time after quench.
Linestyle indicates quenching rates: solid: $\gamma=20$ Gyr$^{-1}$,
dotted: $\gamma=5$ Gyr$^{-1}$,
dashed: $\gamma=2$ Gyr$^{-1}$,
dot-dash: $\gamma=1$ Gyr$^{-1}$,
dot-dot-dot-dash: $\gamma=0.5$ Gyr$^{-1}$,
\label{fig_nmr_quenched}}
\end{figure}

The evolution of NUV-r for the quenched models is graphed in Figure \ref{fig_nmr_quenched},
for a selection of quench rates: $\gamma=[20,5,2,1,0.5]$ Gyr$^{-1}$, or
$\tau=1/\gamma=[0.05,0.2,1,2]$ Gyr.  The transitional color NUV-r=4.25 is reached at different times after the quench time $t_q$,
increasing with $1/\gamma$. 

\begin{figure}
\plottwo{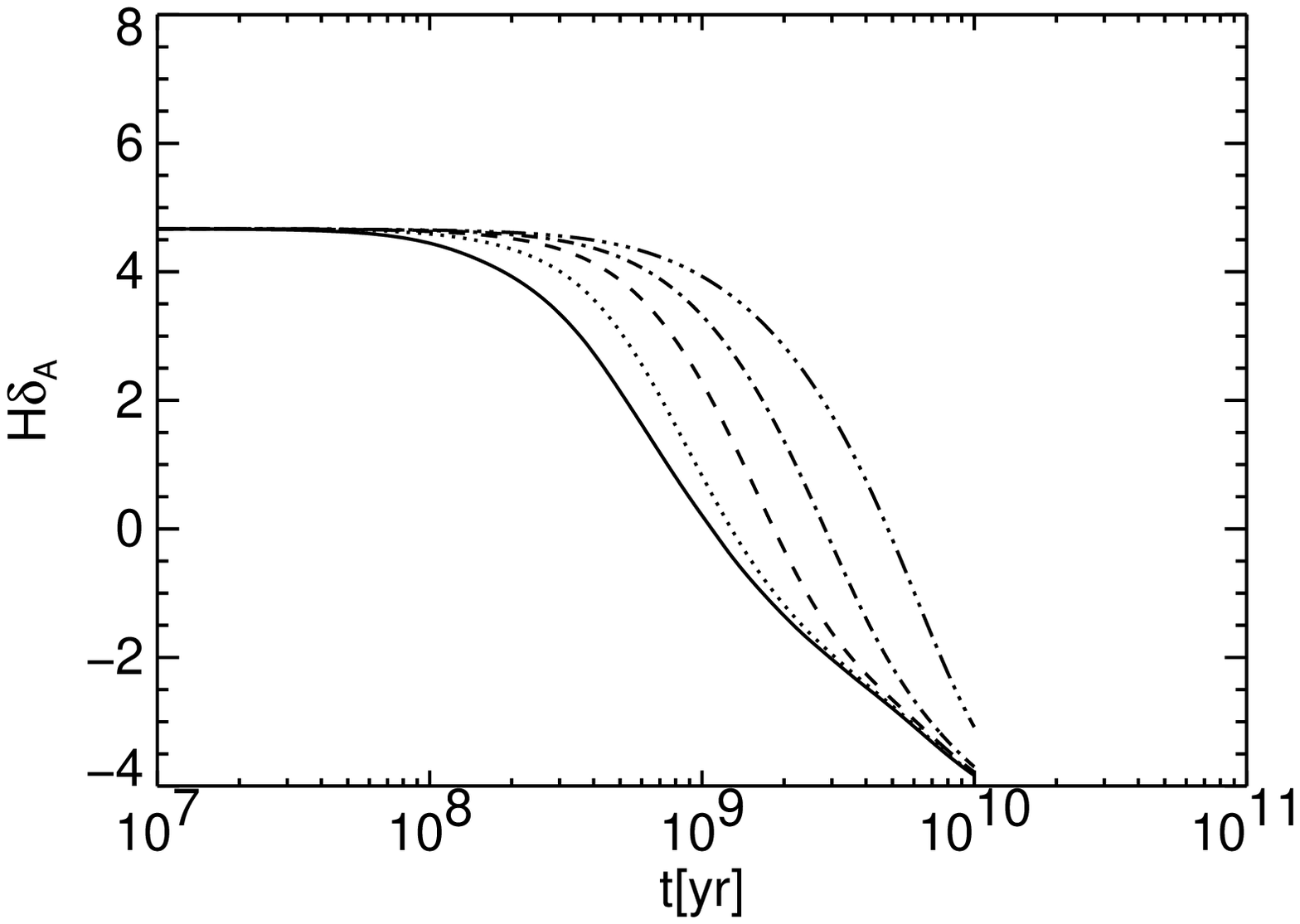}{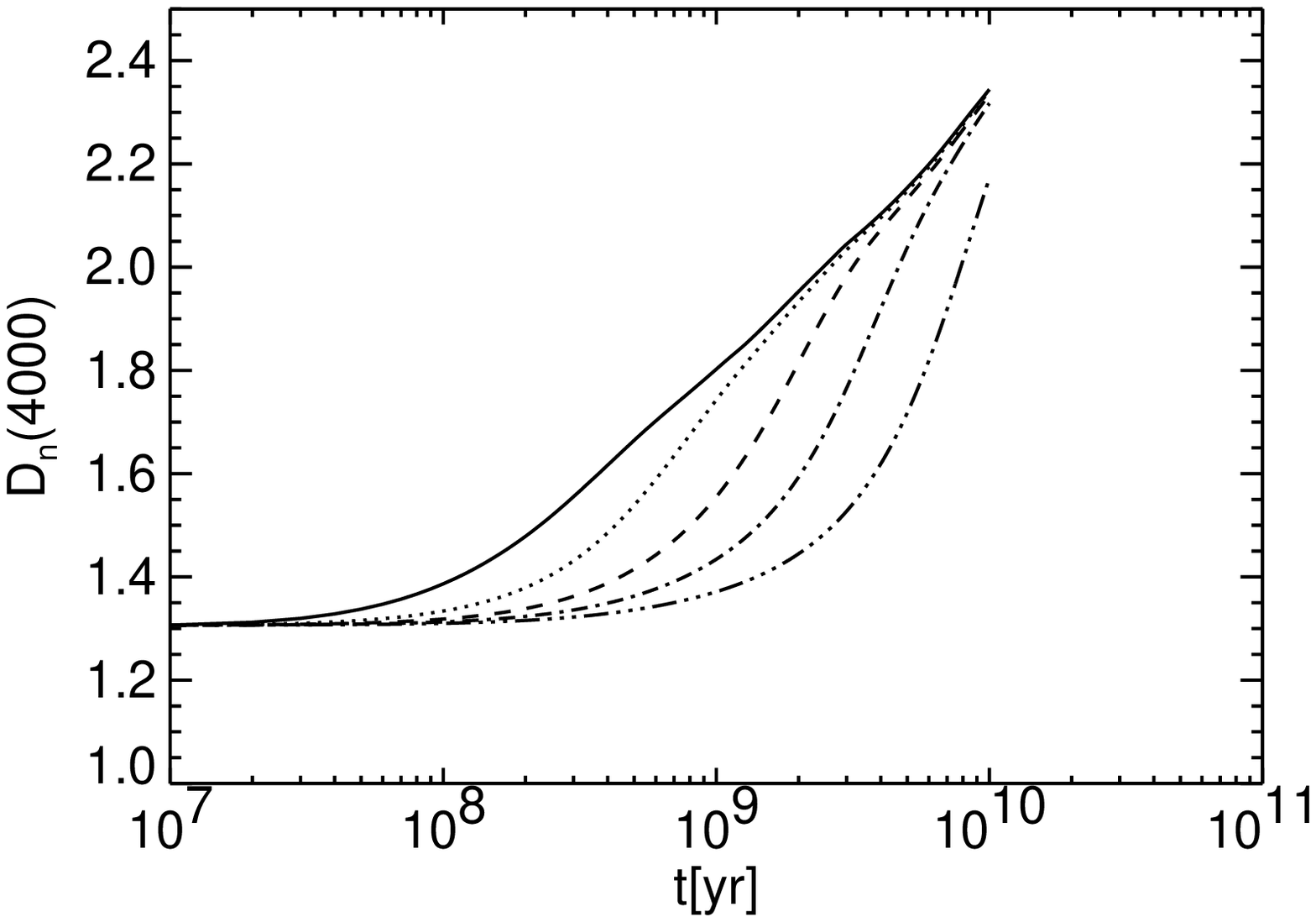}
\caption{Quenched star formation history, \hda [LEFT] and \dn [RIGHT] vs. time after quench.
Linestyle indicates quenching rates: solid: $\gamma=20$ Gyr$^{-1}$,
dotted: $\gamma=5$ Gyr$^{-1}$,
dashed: $\gamma=2$ Gyr$^{-1}$,
dot-dash: $\gamma=1$ Gyr$^{-1}$,
dot-dot-dot-dash: $\gamma=0.5$ Gyr$^{-1}$,
\label{fig_hda_quenched}}
\end{figure}

\begin{figure}
\plottwo{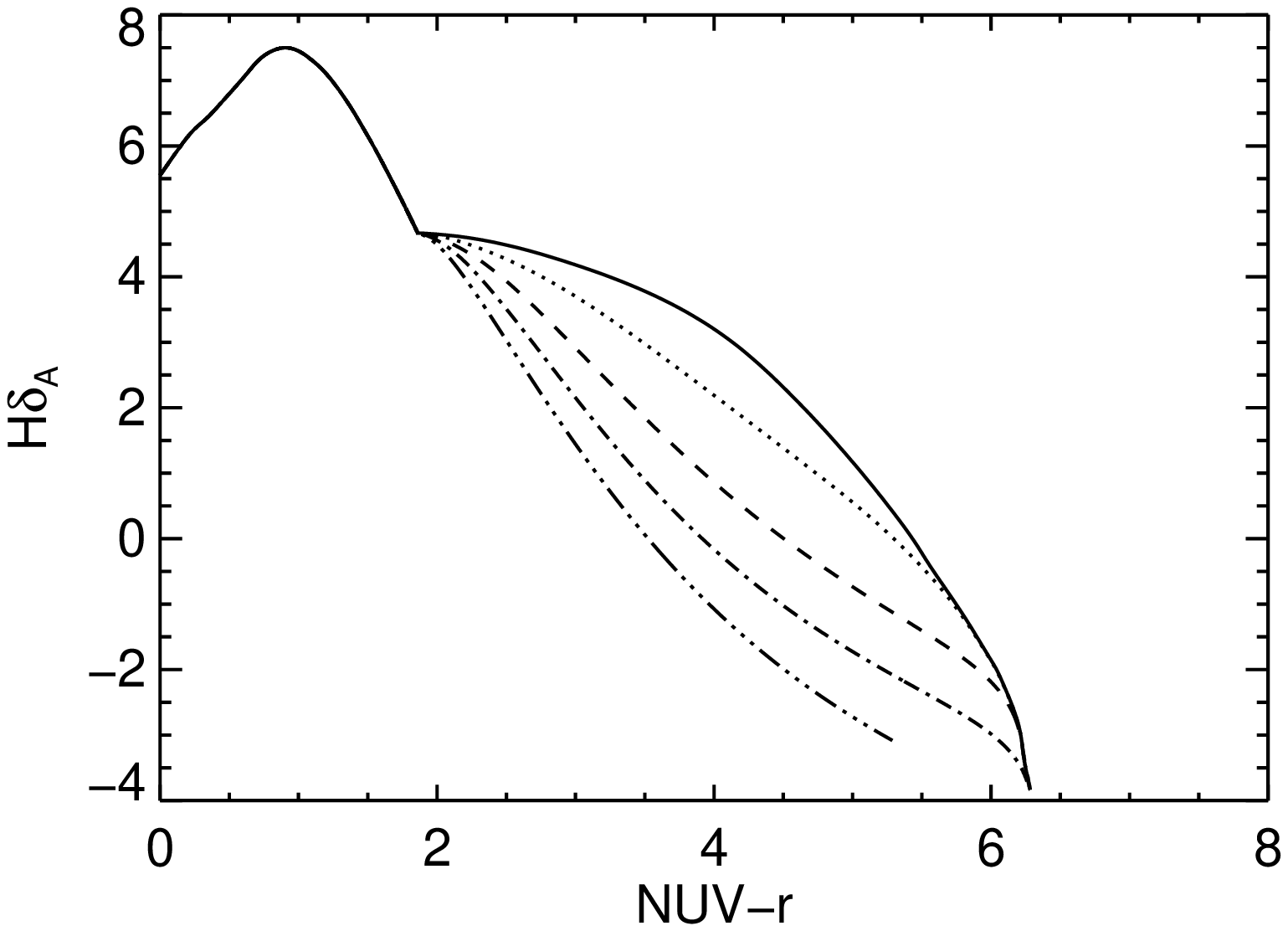}{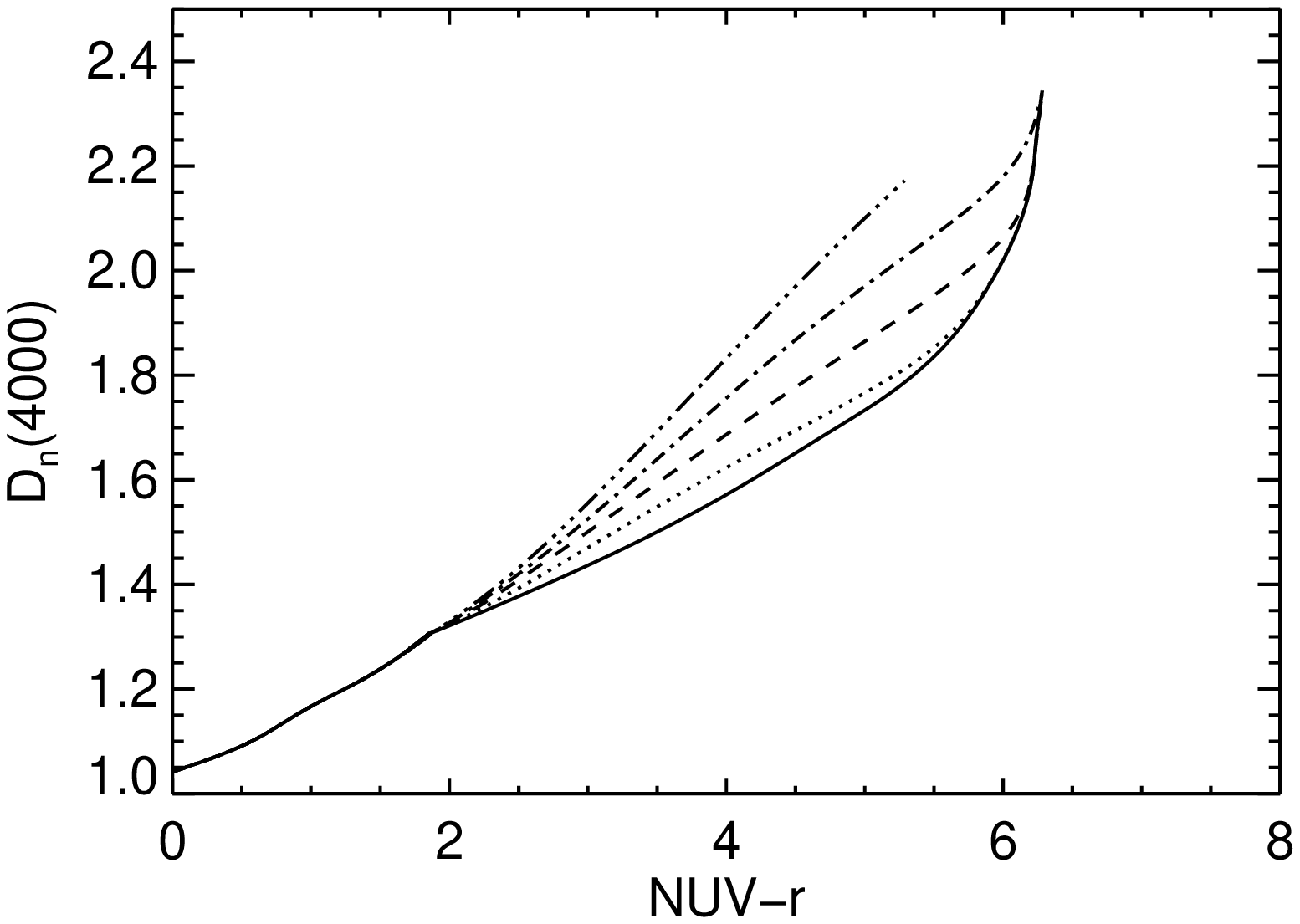}
\caption{Quenched star formation history, \hda [LEFT] and \dn [RIGHT] vs. NUV-r after quench.
Linestyle indicates quenching rates: solid: $\gamma=20$ Gyr$^{-1}$,
dotted: $\gamma=5$ Gyr$^{-1}$,
dashed: $\gamma=2$ Gyr$^{-1}$,
dot-dash: $\gamma=1$ Gyr$^{-1}$,
dot-dot-dot-dash: $\gamma=0.5$ Gyr$^{-1}$,
\label{fig_hda_nmr}}
\end{figure}

In order to calculate the indices \hda and \dn, we use the instantaneous
stellar burst index evolution reported by \cite{kauffmann03} and compute the continuum weighted average
for each star formation history. We find that the indices calculated with the \cite{pickles98}
stellar library are a better fit to the observed indices at later ages (particularly \hda), so we use these. 
The evolution of \hda and \dn for the quenched models is shown in Figures \ref{fig_hda_quenched}, 
\ref{fig_hda_nmr}. 
This illustrates that for our assumed star formation history, and a fixed NUV-r color, 
\hda and \dn are measures of the quench rate $\gamma$. Faster quench rates
yield younger characterisitic stellar ages from \hda (more positive) and \dn (less positive), at a fixed NUV-r color.

\begin{figure}
\plotone{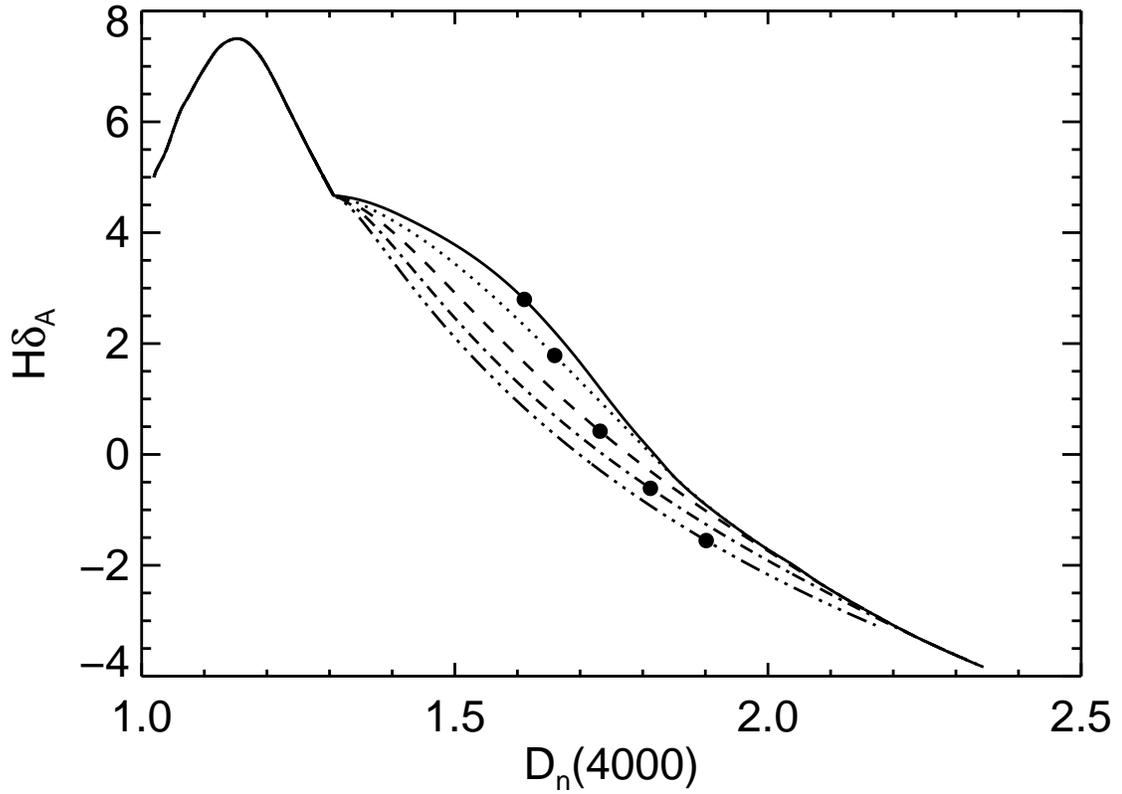}
\caption{Quenched star formation history, \hda vs. \dn after quench.
Linestyle indicates quenching rates: solid: $\gamma=20$ Gyr$^{-1}$,
dotted: $\gamma=5$ Gyr$^{-1}$,
dashed: $\gamma=2$ Gyr$^{-1}$,
dot-dash: $\gamma=1$ Gyr$^{-1}$,
dot-dot-dot-dash: $\gamma=0.5$ Gyr$^{-1}$.
Large points indicate NUV-r=4.25 for each $\gamma$.
\label{fig_hda_d4000}}
\end{figure}

\begin{figure}
\plottwo{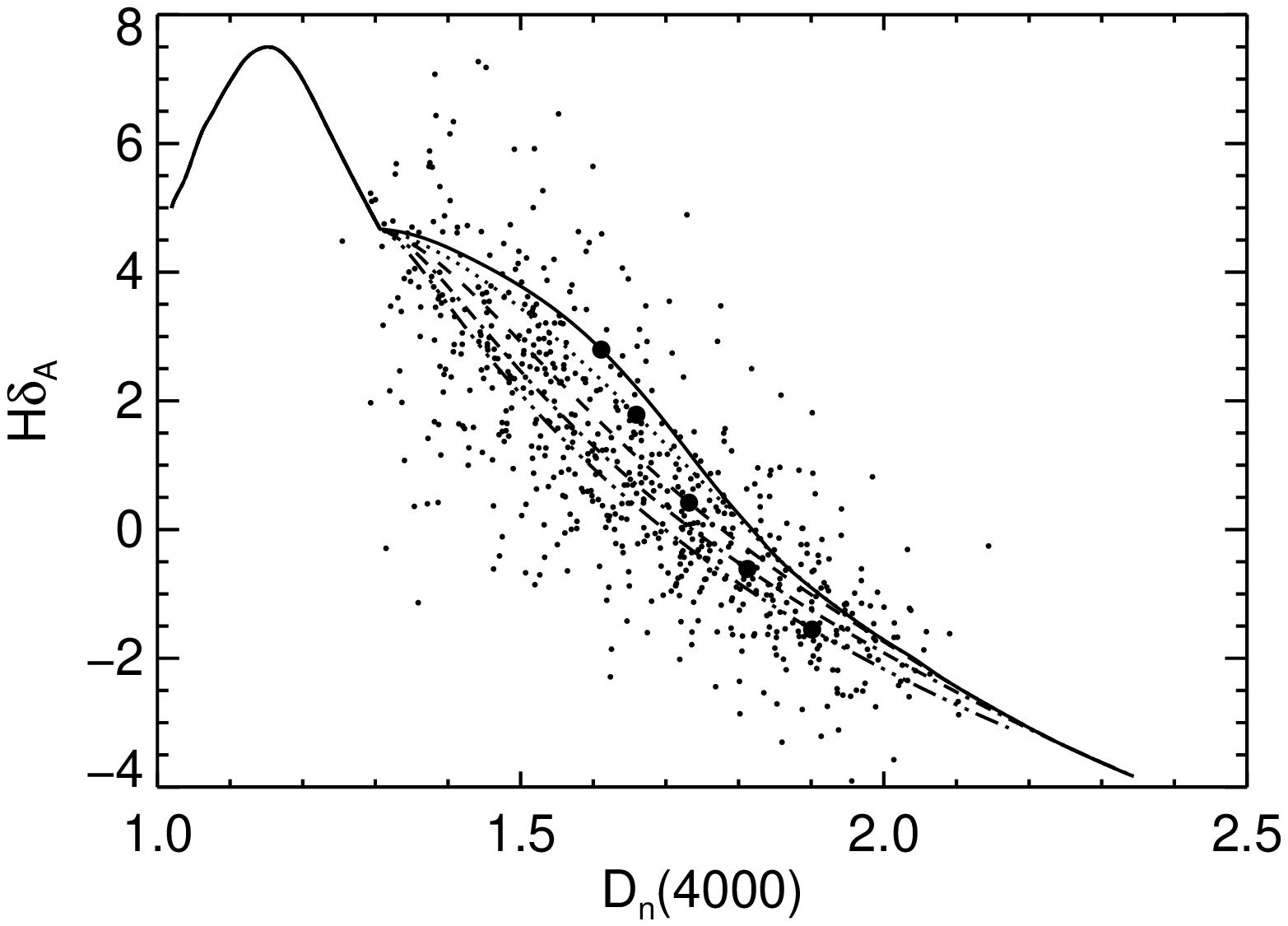}{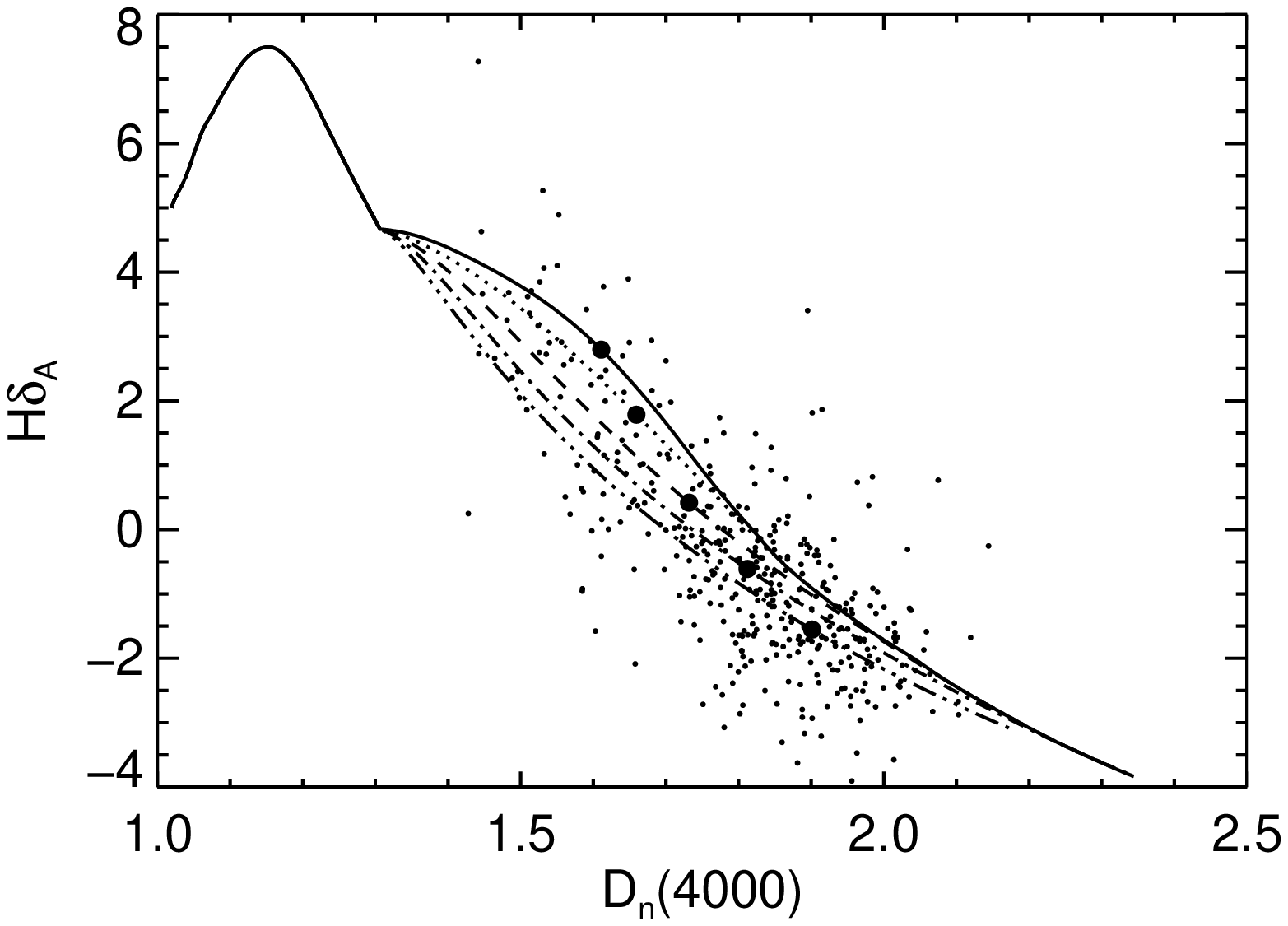}
\caption{Quenched star formation history, \hda vs. \dn after quench.
Linestyle indicates quenching rates: solid: $\gamma=20$ Gyr$^{-1}$,
dotted: $\gamma=5$ Gyr$^{-1}$,
dashed: $\gamma=2$ Gyr$^{-1}$,
dot-dash: $\gamma=1$ Gyr$^{-1}$,
dot-dot-dot-dash: $\gamma=0.5$ Gyr$^{-1}$.
Large points indicate NUV-r=4.25 for each $\gamma$.
Dots show data for a single color-magnitude bin
(M$_r$=-21, NUV-r=4.25). LEFT: no extinction correction. RIGHT: extinction corrected.
\label{fig_hda_d4000_data}}
\end{figure}

In Figure \ref{fig_hda_d4000} we show the evolution of \hda vs. \dn for the various quench rates, and
the points on the locus when NUV-r=4.25. In Figure \ref{fig_hda_d4000_data}, we add the
galaxies from the extinction-corrected \hess~ bin 4$<$NUV-r$<$4.5 and $-22<M_r<-21$ (from Figure \ref{fig_hda_vs_d4000_dat}).
Again, we have made no signal-to-noise cuts. It can be seen that the model indices are a reasonable
representation of the index data and that a range of $\gamma$ can potentially account for much of the
dispersion in a given color-magnitude bin. 

\begin{figure}
\plotone{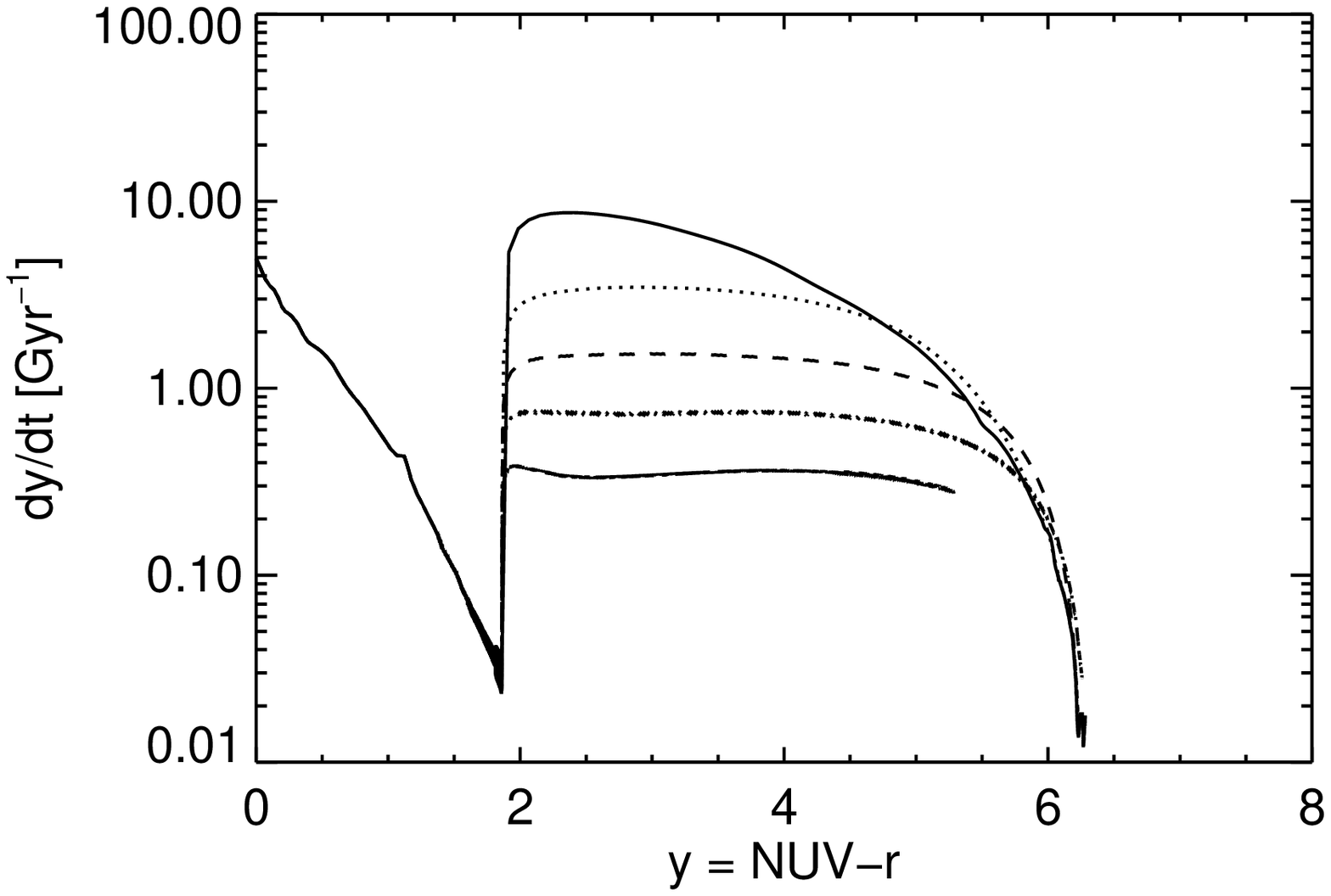}
\caption{Color derivative dy/dt for quenched star formation history vs. NUV-r after quench.
Linestyle indicates quenching rates: solid: $\gamma=20$ Gyr$^{-1}$,
dotted: $\gamma=5$ Gyr$^{-1}$,
dashed: $\gamma=2$ Gyr$^{-1}$,
dot-dash: $\gamma=1$ Gyr$^{-1}$,
dot-dot-dot-dash: $\gamma=0.5$ Gyr$^{-1}$,
\label{fig_dydt_quenched}}
\end{figure}

Finally, in Figure \ref{fig_dydt_quenched}, we show the color derivative $dy/dt$ vs. color for the
quenched models. At $y=4.25$, the derivative varies from a high of $~\sim$ 3 Gyr$^{-1}$ to
about 0.4 Gyr$^{-1}$ for $\gamma=0.5$ Gyr$^{-1}$. Not suprisingly, $dy/dt(4.25)$ monotonically tracks $\gamma$.

We remind the reader that in the spirit of an exploratory investigation
we are using very simple star formation history
models to track the evolution in \dn, \hda, and NUV-r space. We are
also assuming solar metallicity. The star formation history assumption
is further investigated below. We defer a more complex and
realistic approach to the future. 

\section{Constraints on the Mass Flux}

The machinery is now in place to calculate the transitional mass flux.
For all calculations we use the color bin $4<y<4.5$.
We do this in three ways: 

\begin{enumerate} 

\item Use the uncorrected \hess,  calculate a mean
value of \hda and \dn for each $M_r$ bin, infer the corresponding $dy/dt$ from Figures \ref{fig_hda_d4000_data} and
\ref{fig_dydt_quenched}, and use Equation \ref{eqn_sum} to add up the total mass flux.

\item Use the same methodology as 1. but use the extinction-corrected \hess.

\item Calculate a $\gamma$ and $dy/dt$ for each galaxy, and
an average $dy/dt$ ror each color-magnitude bin.

\end{enumerate}

\subsection{Method 1 and 2}

In Tables \ref{tab_method1} and \ref{tab_method2} we tabulate the values of $\phi(M_r, y=4.25)$, $\overline{H\delta_A}(M_r, y=4.25)$,
$\overline{D_n(4000)}(M_r, y=4.25)$, $\gamma$, $dy/dt$, stellar mass (from \cite{kauffmann03} available in the MPIA/JHU Value-Added Catalog), and mass-flux for the uncorrected
and extinction-corrected cases. The quench rate $\gamma$ (and the corresponding
$dy/dt$) are found by finding the y=4.25 (\dn, \hda) model pair that comes
closest to the observed (\dn, \hda) pair. Distance in the (\dn, \hda)
plane is calculated by normalizing each index to its full range, viz. \hda$_n = $\hda/12, and
\dn [norm] = \dn /1.0. 

In the uncorrected case, the total mass flux is $\dot{\rho}_T=0.095$ M$_\odot$ yr$^{-1}$ Mpc$^{-3}$
(the sum of the right-hand column).
The majority of the mass flux comes from galaxies with $-22<M_r<-20$. Lower mass galaxies
have younger stellar ages, faster quench rates and larger $dy/dt$. However, their contribution
to the total mass is decreased. Note that each mass flux entry in the table is obtained from
$\dot{\rho}_T= 2 M(M_r)\phi(M_r,y=4.25) dy/dt[\gamma(\overline{D_n(4000)}, \overline{H\delta_A})]$, where the
factor of two accounts for the fact that the \hess~ NUV-r bins are 0.5 wide. 

In the extinction corrected case, the total mass flux is $\dot{\rho}_T=0.028$ M$_\odot$ yr$^{-1}$ Mpc$^{-3}$,
a factor of two lower. Two effects cause this: the number density $\phi$ is lower (since some star forming
galaxies are moved to a bluer NUV-r bin, and the characteristic ages are somewhat older (for the same reason). 
The distribution shifts to higher $M_r$ bins, with the major contribution coming
from $-23<M_r<-20.5$.  This second method is likely to be applied in deep galaxy samples
for example using stacked spectra to study the mass flux evolution.

\begin{figure}
\plotone{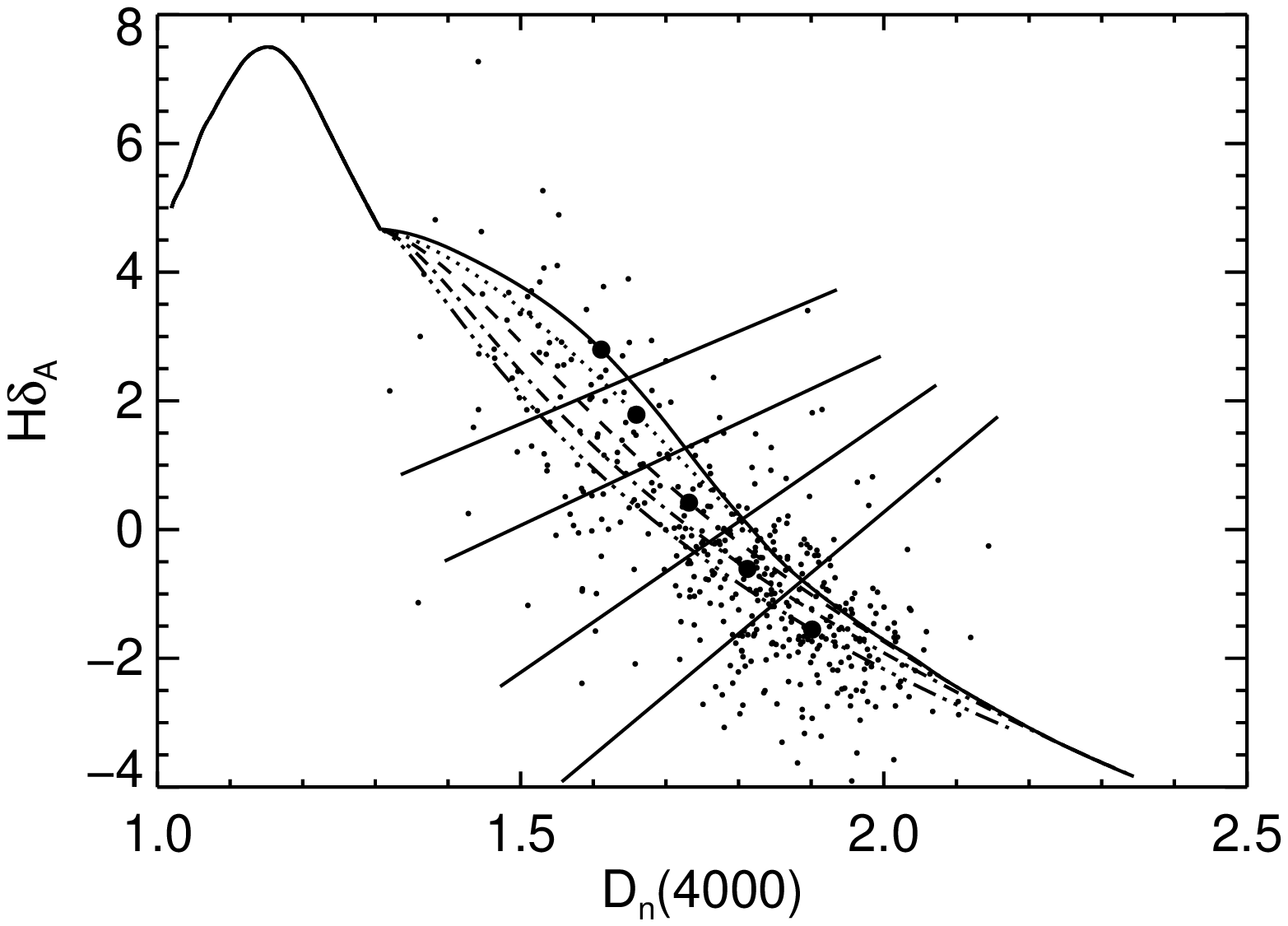}
\caption{Quenched star formation history, \hda vs. \dn after quench.
Linestyle indicates quenching rates: solid: $\gamma=20$ Gyr$^{-1}$,
dotted: $\gamma=5$ Gyr$^{-1}$,
dashed: $\gamma=2$ Gyr$^{-1}$,
dot-dash: $\gamma=1$ Gyr$^{-1}$,
dot-dot-dot-dash: $\gamma=0.5$ Gyr$^{-1}$.
Large points indicate NUV-r=4.25 for each $\gamma$.
Dots show data for a single color-magnitude bin
(M$_r$=-21, NUV-r=4.25) for the extinction corrected case.
Straight lines indicate bins used to generate Figure \ref{fig_method3_bin}.
\label{fig_hda_d4000_bin}}
\end{figure}

It is interesting to examine the distribution of galaxies vs. $\gamma$.
We have binned the galaxies into five $\gamma$ bins, shown in Figure \ref{fig_hda_d4000_bin}.
In Figure \ref{fig_method3_bin} we show the color derivatives $dy/dt$, the
fraction of galaxies in each bin, and the fraction weighted by the color derivative $dy/dt$
which compensates for the briefer residence time of fast quench objects.
While the unweighted distribution shows a preponderence of 
slow-quench galaxies, the weighted distribution shows that the underlying distribution is fairly uniform.
Galaxies with faster quench rates will spend less time in the transition
region, and the fact that the majority of galaxies are found to have low quench
rates in the transition zone does not mean that the majority of galaxies
undergo slow quenches. 

\subsection{Method 3}

Since there is considerable variance in the age indicators in a given color-magnitude bin,
and the non-linear dependence of $dy/dt$ on the various age indices,
we calculate the $\gamma$ and $dy/dt$ for each galaxy, and then determine the
average $dy/dt$ and its error due to the measurement error of \dn and \hda, within a particular color-magnitude bin. 
We perform this average for each $M_r$ bin, and display the results in Table \ref{tab_method3}.
The color derivatives and resulting mass fluxes are $\sim$15\% larger. The resulting summed mass flux is
$\dot{\rho}_T=0.033\pm 0.001$ M$_\odot$ yr$^{-1}$ Mpc$^{-3}$. The small reported error is the
formal error due to the Poisson error in the \hess and the index measurement error. The
errors in the individual $M_r$ bins are given in Table \ref{tab_method3}, and are typically $\sim 10-25$\%. 

\subsection{Comparison to Mass Growth of the Red Sequence}

Two recent major studies, \cite{bell04b} using COMBO-17, and \cite{faber05} using DEEPII and COMBO-17, have
concluded that the red sequence has grown significantly over the period z=1 to z=0. 
In particular, \cite{faber05} measured an increase in the luminosity function
normalization of $\Delta\phi_0 = 0.56\pm0.09$ dex since $z=1.0$ and
$\Delta\phi_0 = 0.36\pm0.09$ dex since $z=0.8$. 

We convert this to a mass flux using the following relationship:
\begin{equation}\label{eqn_redflux}
\dot{\rho}_R \simeq j_B(RS) \left( {M \over L_B } \right) \left( {{10^{\Delta\log{\phi_0}} - 1} \over {10^{\Delta\log{\phi_0}}}} \right) {1 \over {t(z=0.1)-t(z)}}
\end{equation}
\cite{faber05} estimate that the mass-to-light ratio ${M \over L_B }=6$ for today's red-sequence galaxies, and the red sequence B-luminosity density at $z=0$ is
$j_B=10^{7.7}$ L$_\odot$ yr$^{-1}$ Mpc$^{-3}$. This gives a mass flux of $\dot{\rho}_R=0.034 (0.031)$  M$_\odot$ yr$^{-1}$ Mpc$^{-3}$ for
$0<z<1$ ($0<z<0.8$), both with a minimum error of $\pm$0.1 dex. By doing this we assume that the mass
flux is constant over the redshift interval considered, in order to compare it to the blue sequence mass
flux obtained ``today'' (where for today we use z=0.1). 

\begin{figure}
\plotone{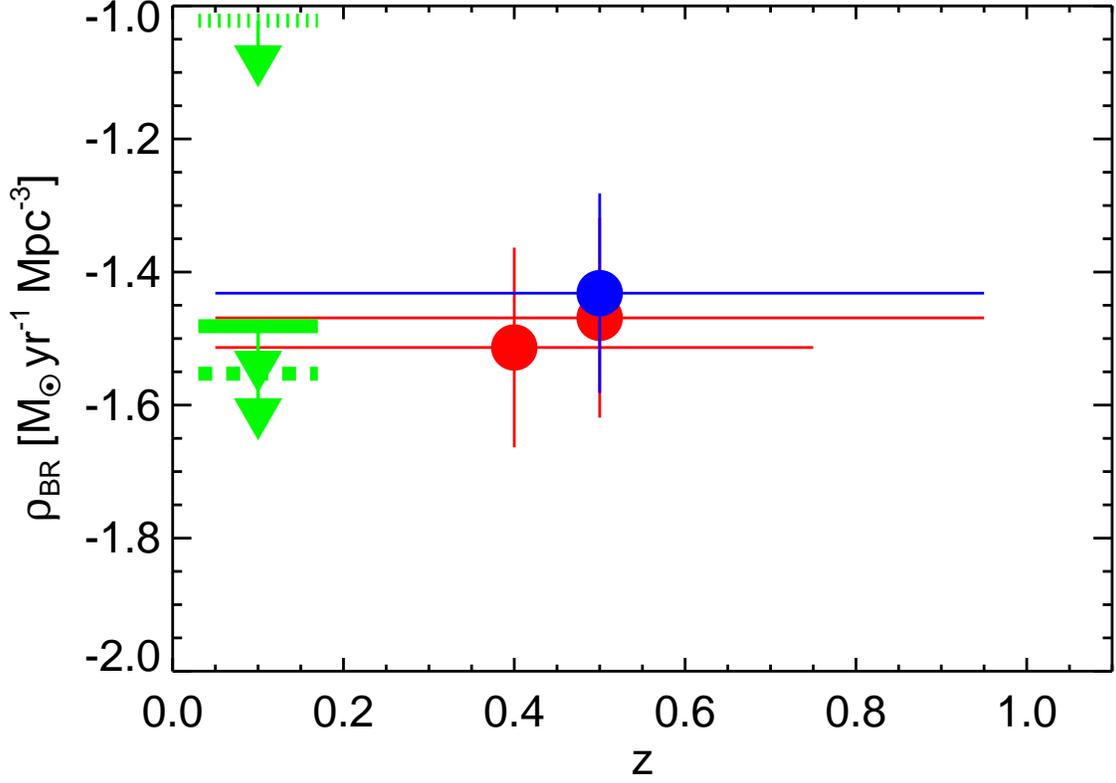}
\caption{Results from this work for the transitional mass flux from the blue to
red sequence. Green dashed line shows result of method 1, green dotted line
method 2, green solid line method 3. Red points show mass flux estimated from
red sequence evolution of \cite{faber05}. Higher point is based on
evolution over $0<z<1$, while lower point on $0<z<0.8$. Blue point
shows estimate $-\dot{\rho}_B$ based on blue sequence evolution. 
\label{fig_rho}}
\end{figure}

We compare these results to the derived blue sequence mass fluxes using Method 1 and Method 3 in Figure \ref{fig_rho}.
The horizontal bars indicate the redshift range over which the flux has been averaged.
For reasons that we discuss in the next section, we consider the blue flux measurement as an upper limit.
It is intriguing that the blue sequence and red sequence mass fluxes are in rough agreement. 
This comparison makes sense only if we are measuring mass fluxes at comparable epochs (in general we are not)
or if the mass flux is constant over time (our principal assumption). The overall conclusions of \cite{faber05} are not inconsistent with
a constant red sequence mass growth rate over $0<z<1$, and although there is some evidence
for a slightly increased rate for $0.8<z<1$, the average rate obtained over both redshift intervals
is within 10\%.  We return to this issue in the section \ref{sec_discussion}.

\subsection{Comparison to Mass Loss from the Blue Sequence}

To calculate the blue sequence mass flux, we can use the change in the blue luminosity density, but
we also need to account for evolution in the mass-to-light ratio. Blanton \cite{blanton06}
has shown by comparing SDSS and DEEP2 luminosity and color-magnitude distributions
that the blue sequence becomes redder by $(u-g)_{0.1}[z=0.1]-(u-g)_{1.0}[z=1]\sim0.3$
and luminosity decreases by $\sim 1$ mag. This is consistent with an increase mass-to-light ratio
from $(M/L_B)[z=1]\simeq1.3$ to $(M/L_B)[z=0.1]=2.5$ \cite{bell03}.  Then, with
\begin{equation}
\dot{\rho}_B \simeq {{j_B(BS) \left( {M \over L_B } \right)[z=1] - j_B(BS) \left( {M \over L_B } \right)[z=0.1]} \over {t(z=0.1)-t(z)}}
\end{equation}
and $j_B(BS)[z=1]=10^{8.2}$ and $j_B(BS)[z=0.1]=10^{7.8}$,  for blue-sequence galaxies gives
$-\dot{\rho}_B\simeq0.01$  M$_\odot$ yr$^{-1}$ Mpc$^{-3}$. 
This mass flux is low, as Blanton \cite{blanton06} concluded, at least a factor of
four below our measured transition flux. This conclusion is not
changed by extinction, since the extinction vector lies along the $(M/L_B)$ vs. $u-g$ locus
\citep{bell03}. 

How can galaxies be populating the transition zone without depleting
the mass of the blue sequence? As suggested by the anonymous referee, 
there is  significant mass growth on the blue sequence due
to ongoing star formation. If we integrate the star formation history
derived for example by GALEX measurements \citep{schiminovich05} and then
(very crudely) divide by the age as in
\begin{equation}
\dot{\rho}_{SF} \simeq{ { {\int_{z_0}^{z_1}} \dot{M_*}(z) dz} \over {t(z_0)-t(z_1)}},
\end{equation}
we find that 
$\dot{\rho}_{SF}\simeq0.037\pm0.006 $  M$_\odot$ yr$^{-1}$ Mpc$^{-3}$ for $z_0=0.1$
and $z_1=1.0$, and $\dot{\rho}_{SF}\simeq0.031\pm0.006 $  M$_\odot$ yr$^{-1}$ Mpc$^{-3}$ 
for $z_1=0.8$. 

Remarkably, it appears that on-going star formation balances the 
mass flux off the blue sequence. In fact, mass continuity, the constancy of the
blue sequence mass found by \cite{blanton06}, and
the measured star formation history strongly argue that
galaxies are transitioning off the blue sequence at the rates
derived independently in this paper. It will be interesting to
understand whether this balance is merely a coincidence.

\section{Discussion\label{sec_discussion}}

\subsection{Issues and Caveats}

\subsubsection{Aperture Corrections}

SDSS spectroscopy is obtained with 3 arcsecond fibers, corresponding to a physical
diameter of 3 kpc to 10 kpc for $0.05<z<0.2$, where the bulk of our sample lies. A possible concern is that
this leads to aperture bias. In our case, we might worry that disk light is underrepresented or bulge overpresented
in the \hda and \dn indices. This would lead to an underestimate of the quench rate and $dy/dt$.
If this were the case, we would expect that the more distant galaxies in the sample would have higher
\hda (and lower \dn) as the disk is brought into the fiber aperture. However, no such trend is apparent in the data--there
is no redshift dependence at all in the index values.

\subsubsection{Time Variations and Delays} 

In comparing the transitional, red, and blue-sequence mass fluxes, we
have glossed over time delays in the transition from blue to red sequence. These could impact the present
measurement if there are variations in the mass flux rate over the cosmic timescales spanned by the 
DEEPII and COMBO-17 observations.
We have determined typical quench rates to be $\sim$1-1.5 Gyr$^{-1}$. With this rate, evolution from
blue to our transitional color takes $\sim 3$ Gyr, with a similar additional interval to reach
the red sequence. Strictly speaking, we must compare the mass flux rate now with the blue sequence evolution
3 Gyr in the past and the red sequence evolution 3 Gyr in the future. The results of \cite{faber05} suggest
that the mass flux rate is roughly constant, although the rate may be higher in the highest
redshift bin. It will be of some interest to measure the transitional
flux at higher redshifts, to determine whether it is evolving with time. 

\subsubsection{Star Formation Histories} 

We have assumed that all galaxies in the transitional region follow the quenched star formation history of
equation \ref{eqn_qsfh}. Many other star formation histories could be considered which would
produce different evolution in \hda, \dn, color space. We maintain, however, that
the quenched models we use lead to the highest possible color derivatives, $dy/dt$, in the transitional region.
In this sense, since the mass flux we derive is proportional to $dy/dt$, our result represents an upper limit.

We assumed that the quench occurs after 10 Gyr of constant star formation. This
produces some galaxies (with the oldest \dn and \hda indices) that exceed the
age of the universe by several Gyrs. If the constant SFR period
is reduced to 5 Gyr, the implied quench rate for a given index is lower, and correspondingly
the total mass flux is reduced by about 20\% from 0.026 to 0.021 M$_\odot$ yr$^{-1}$ Mpc$^{-3}$ (Method 3). 

We consider three other possible histories, a smooth exponential decay in star formation rate from formation
({\it smooth}), recent bursts, and a composite stellar population, consisting of an early burst and a low level of
on-going star formation ({\it composite}). 

\begin{figure}
\plotone{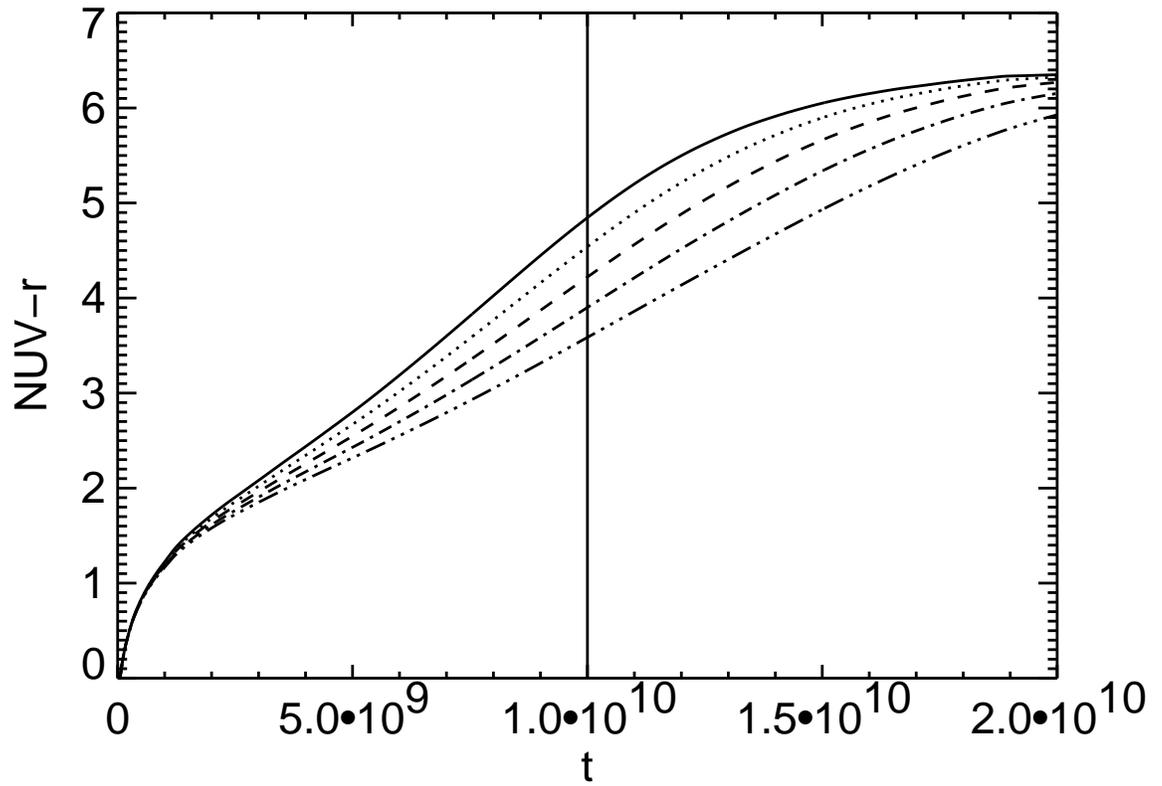}
\caption{Smooth, exponentially decaying star formation rates yield colors NUV-r=4.25
near 10 Gyr ages. Linestyle indicates exponential constant: solid: $\gamma=0.6$ Gyr$^{-1}$,
dotted: $\gamma=0.55$ Gyr$^{-1}$,
dashed: $\gamma=0.5$ Gyr$^{-1}$,
dot-dash: $\gamma=0.45$ Gyr$^{-1}$,
dot-dot-dot-dash: $\gamma=0.4$ Gyr$^{-1}$.
\label{fig_nmr_tsfr_smooth}}
\end{figure}

\begin{figure}
\plottwo{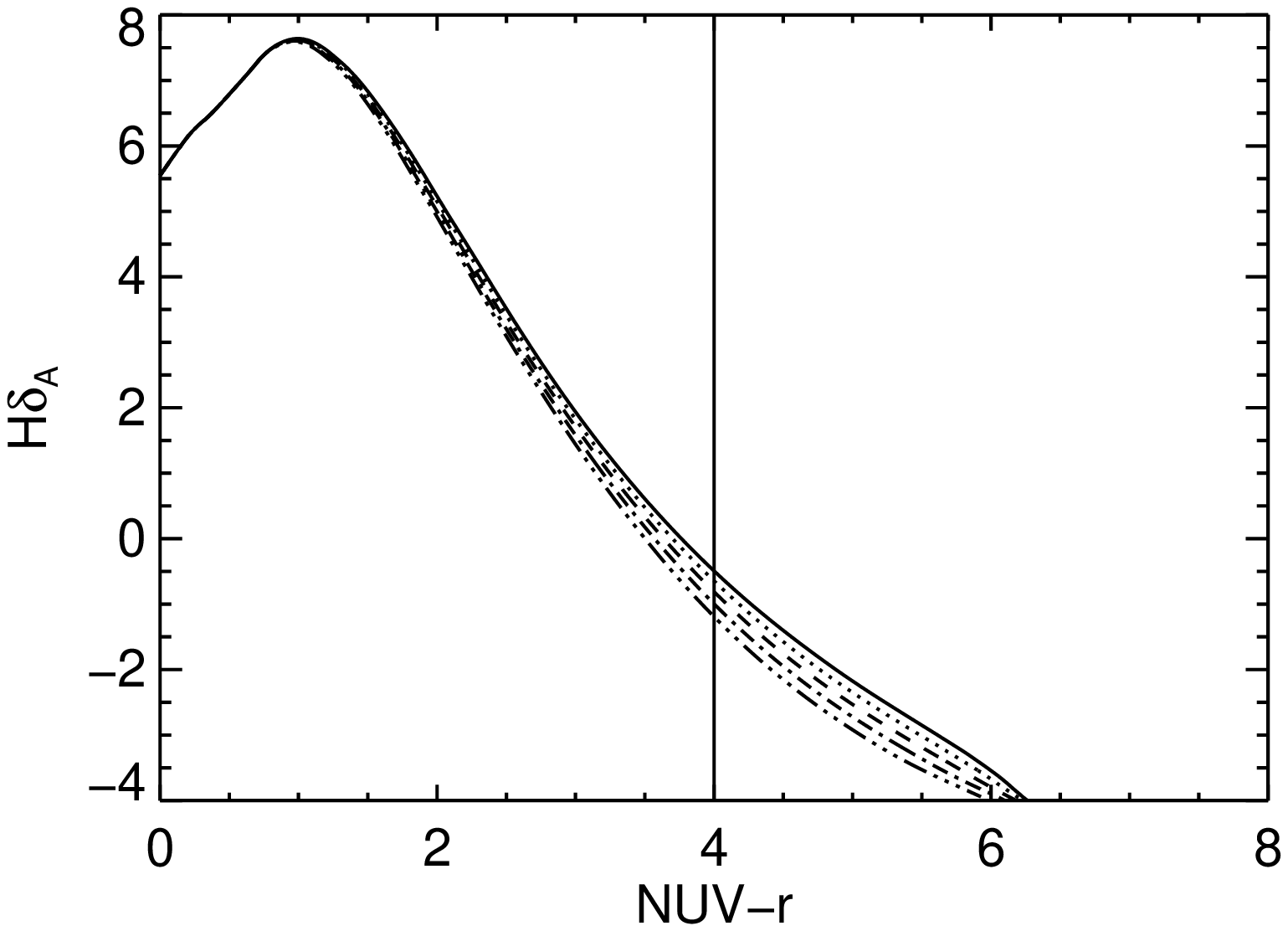}{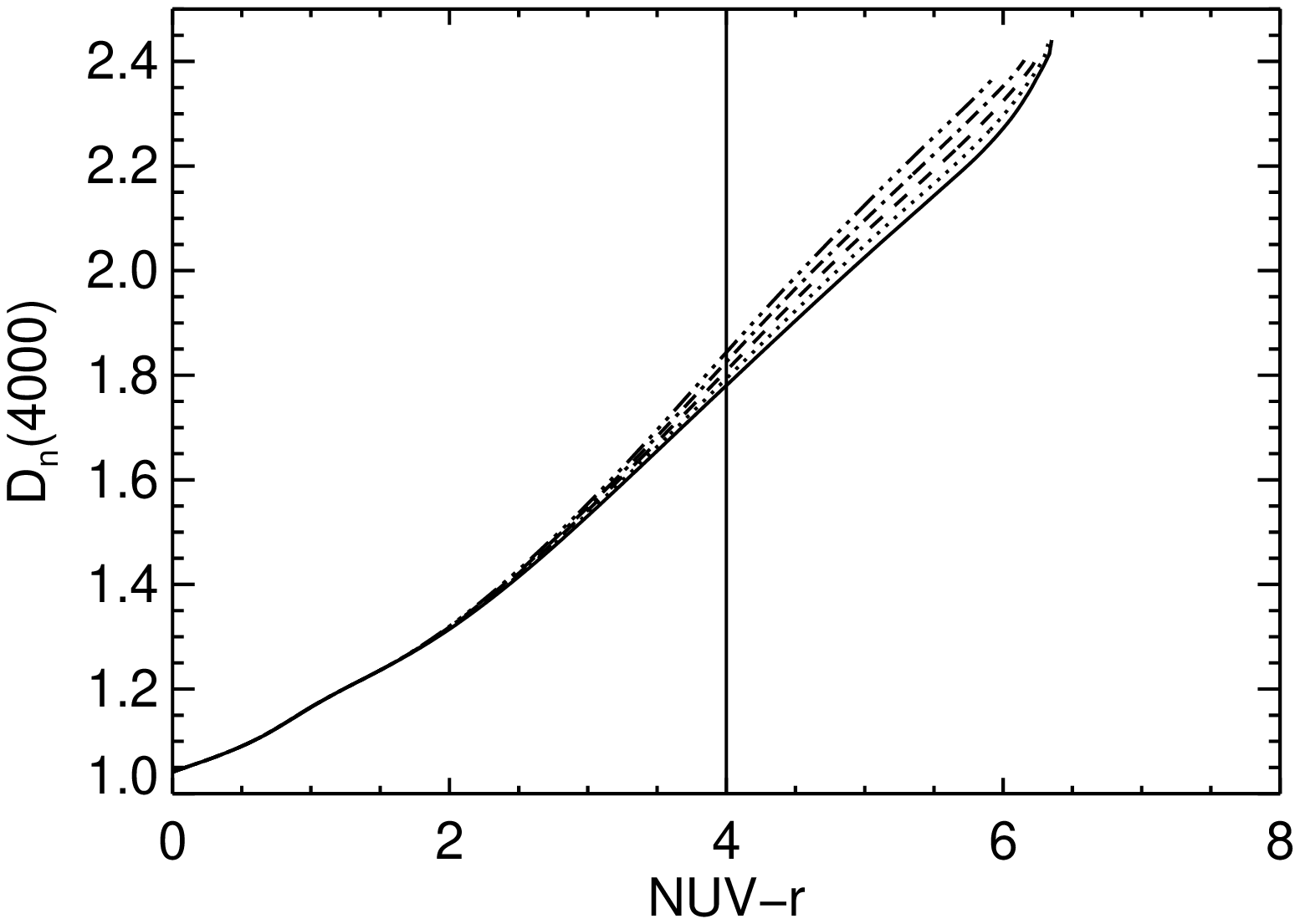}
\caption{Smooth, exponentially decaying star formation rates \hda and \dn indices
vs. NUV-r color.
Linestyle indicates exponential constant: solid: $\gamma=0.6$ Gyr$^{-1}$,
dotted: $\gamma=0.55$ Gyr$^{-1}$,
dashed: $\gamma=0.5$ Gyr$^{-1}$,
dot-dash: $\gamma=0.45$ Gyr$^{-1}$,
dot-dot-dot-dash: $\gamma=0.4$ Gyr$^{-1}$.
\label{fig_index_nmr_smooth}}
\end{figure}

{\it Smooth}. In this case we assume that galaxies were formed $\sim$10 Gyr ago with a range of e-folding parameters $\gamma$
that accounts for the range in present day colors, the original assumption of \cite{tinsley68}. We show the
time evolution of NUV-r in Figure \ref{fig_nmr_tsfr_smooth}. The time at which the
transitional color NUV-r=4.25 is reached depends somewhat upon the value of $\gamma$.
We show the color evolution of \hda and \dn in Figure \ref{fig_index_nmr_smooth}.

\begin{figure}
\plotone{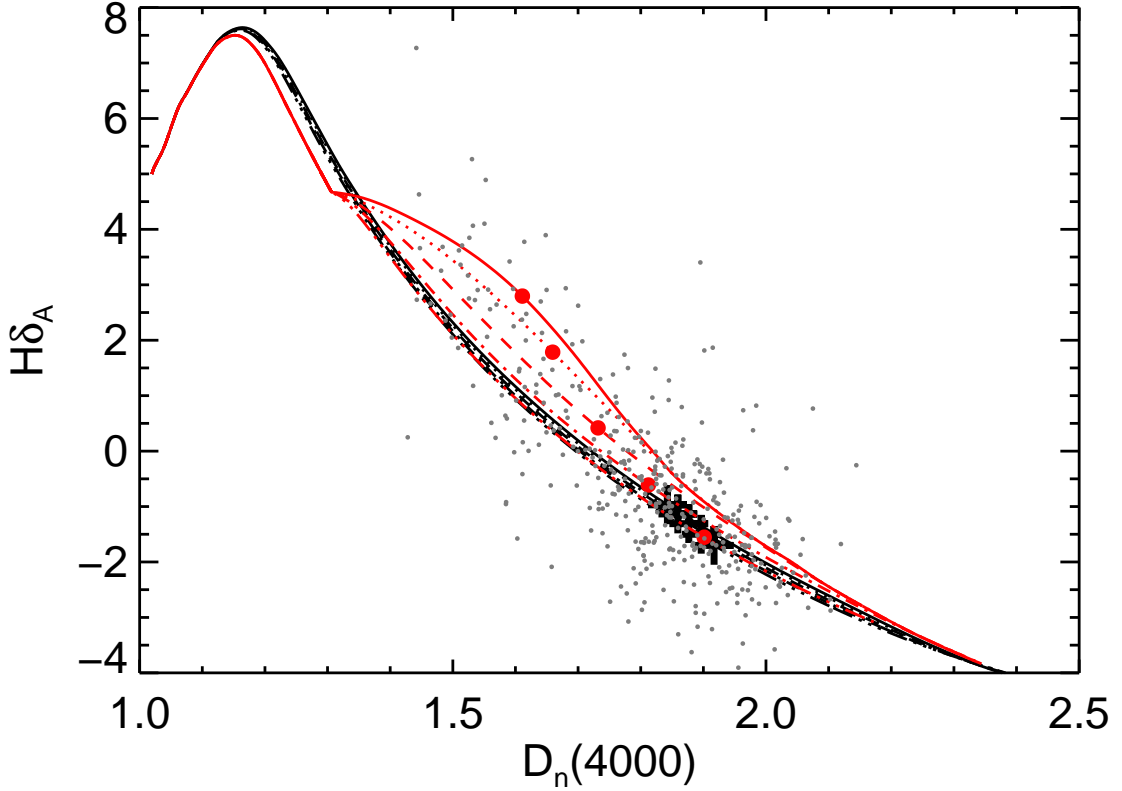}
\caption{Black lines give smooth, exponentially decaying star formation rates \hda vs. \dn 
Linestyle indicates exponential constant: solid: $\gamma=0.6$ Gyr$^{-1}$,
dotted: $\gamma=0.55$ Gyr$^{-1}$,
dashed: $\gamma=0.5$ Gyr$^{-1}$,
dot-dash: $\gamma=0.45$ Gyr$^{-1}$,
dot-dot-dot-dash: $\gamma=0.4$ Gyr$^{-1}$.
Large crosses indicate epoch when NUV-r=4.25.
In red, quenched models are plotted with the following linestyles:
solid: $\gamma=20$ Gyr$^{-1}$,
dotted: $\gamma=5$ Gyr$^{-1}$,
dashed: $\gamma=2$ Gyr$^{-1}$,
dot-dash: $\gamma=1$ Gyr$^{-1}$,
dot-dot-dot-dash: $\gamma=0.5$ Gyr$^{-1}$.
Large points indicate epoch when NUV-r=4.25 for each $\gamma$.
Grey dots are data for $-22<M_r<-20$. 
\label{fig_smooth_dat}}
\end{figure}

We show a summary index comparison of the smooth models, quenched models, and data
in Figure \ref{fig_smooth_dat}.
The smooth models follow roughly the same locus as the slowest quench rate curve 
($\gamma=0.5$ Gyr$^{-1}$ for the quenched model.
The points on the locus where NUV-r reaches 4.25 is shown for both model sets. 
Smooth models reach NUV-r=4.25 at ``old'' index values,
corresponding again to the index values at NUV-r=4.25 for the $\gamma=0.5$  Gyr$^{-1}$ quenched model.
The figure also shows the index data
(for galaxies in $M_r$ bins which contribute the bulk of the mass flux derived
above, $-22<M_r<-20$). The preponderance
of galaxies falls at higher \hda and lower \dn (above and to the left) than predicted by the smooth models
at the transitional color. Many of the galaxies fall above the smooth \hda-\dn locus at all colors,
where the quenched models provide a better match to both indices at the correct color.
For the small fraction of the galaxies have indices and colors consistent with the smooth
models, the corresponding color derivative $dy/dt\simeq 0.3$ is very close to that
for the low quench rate model with the same $\gamma$. In other words, for long decay times, 
the prior star formation history does not affect the colors and indices: smooth and quenched
models give similar results, and our methodology still applies. Clearly, however, the
majority of galaxies are not well represented in the index-color diagram with
smoothly declining models of age $\sim$10 Gyr. 

Smooth models with significantly higher
values of $\gamma$ can produce the transitional colors but only
if galaxies continue to form at late times. These models can reproduce the
younger spectroscopic ages as well, and actually lead to color derivatives
and mass fluxes that are not dissimilar to those derived for quenched models.
Late formation (rather than late dry merging) of a 
significant population of red sequence galaxies does not
however accord with most observations. 

{\it Bursts.} A massive red-sequence galaxy suffering a new starburst due to a minor merger
with a gas-rich galaxy will evolve from the red to the blue sequence,
and then may evolve back to the red sequence after the gas supply is exhausted
by star formation or feedback.
Some fraction of galaxies in the transitional color bins may be in this state moving
in either direction.  As long as the starburst ramp up time is short, the transition
from red to blue will be very fast, with little impact on the transition
color number densities. However the subsequent decay is determined again by
the ``quench time'' of the starburst. In general the complexity of possible burst histories makes the
present simple treatment inadequate. It can be shown that the quench rate,
\hda and \dn evolution, and color derivative are similar to that of the
quenched case studied above, but with quantitive differences of order 2-3.
There are other subtleties which could influence the 
derivation of a mass flux from the red sequence evolution of \cite{faber05}(such as mass-to-light ratio). 
Regardless, bursting previously red sequence galaxies produce additional
numbers in the transitional colors as they return with considerably lower mass flux. 
We are secure then in interpreting our mass flux measurement as an upper limit.

\begin{figure}
\plotone{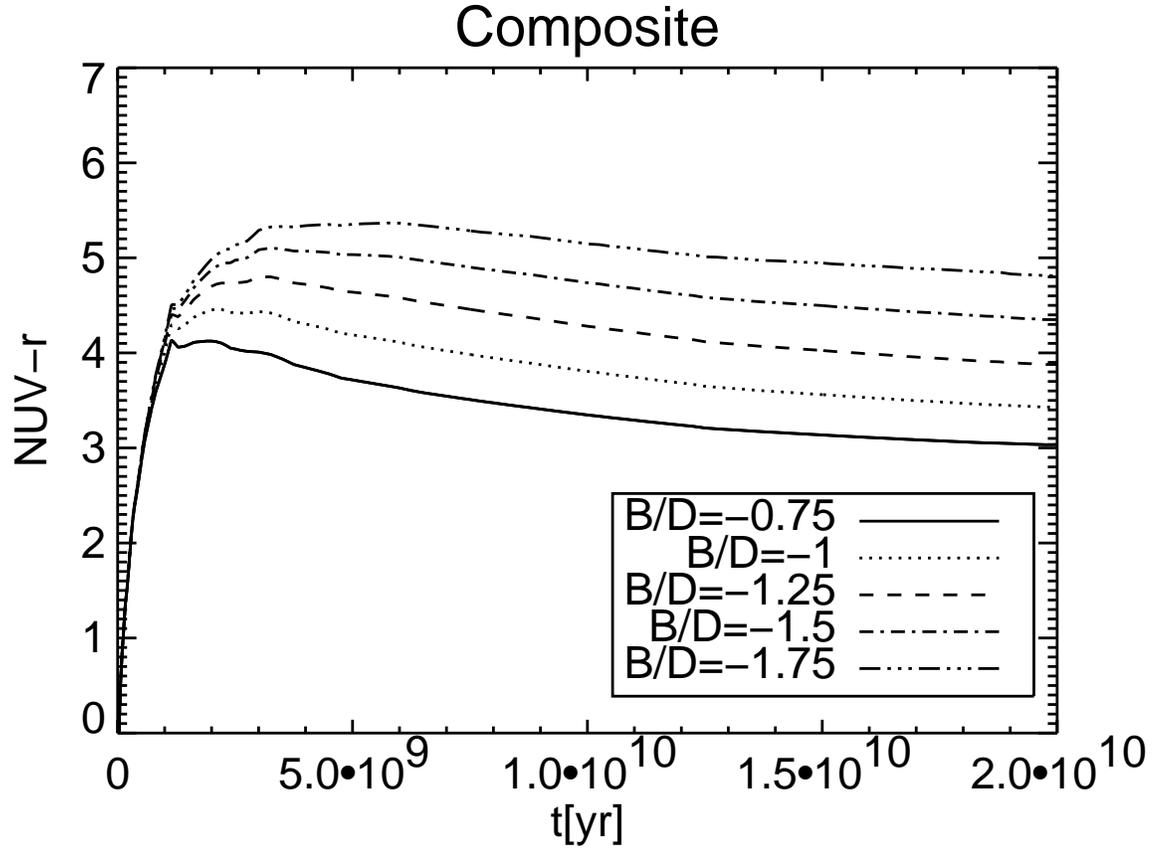}
\caption{NUV-r color vs. time since formation for composite stellar population model with final bulge-to-disk ratio in
mass after 20 Gyr given by the legend. The composite model combines a 10$^{11}$ M$_\odot$ $\tau=0.01$ Gyr exponential bulge
with a constant star formation rate disk. The model with log B/D=-1.25 produces galaxies in the transition color NUV-r=4.25.
\label{fig_nmr_composite}}
\end{figure}

{\it Composite Histories.}
An early burst superimposed with low-level, on-going star formation can produce a galaxy
of any intermediate color if properly tuned. We show this in Figure \ref{fig_nmr_composite}.
For example, a 10$^{11}$ M$_\odot$ bulge can be combined with a
10$^{9.75}$ M$_\odot$ constant star formation rate disk, which corresponds to ongoing star formation that has
produced 3\% of the total mass after 10 Gyr, produces a color which hovers around NUV-r$\simeq$4 after about 10 Gyr.
Objects such as this fall on the lower right of the \hda-\dn diagram:
at the transition color, \hda$=-2.1$ and \dn$=1.96$, rather old values.
In other words the relatively low level of ongoing star formation does significantly perturb
the spectral indices from that of an old population.
The color derivative for this model at the transition color is low and negative, $dy/dt\sim-0.1$.
Certainly, there is evidence for low level star formation in nearby ellipticals \citep{yi05}.
The contribution of galaxies in this bin, as we showed in Figure \ref{fig_method3_bin}, is relatively
small since the color derivative is low anyway. Zeroing this contribution would not signficantly
affect our result. Alternatively, we can simply note that these galaxies have no
impact on our result interpreted as an upper limit.

\subsubsection{Extinction}

We have performed an extinction correction to remove from the transition region
heavily reddened blue-sequence galaxies. This led to about a factor of two
decrease in the derived mass flux. The extinctions were obtained for the z-band
by \cite{kauffmann03} using continuum fitting to the star formation histories.
The NUV light-weighted stellar age in the quenched models is $\sim$1 Gyr.
It is therefore reasonable to use the z-band continuum extinctions in most of these
transition color galaxies because the NUV light is tracing 
stellar populations which have diffused away from star forming regions. 
The \cite{kauffmann03} suite of star formation histories included smooth declining exponentials
plus bursts. Their suite did not explicitly include the kind of quenched models we use,
although the combination of bursts and declining models could mimic them.
We examined the sensitivity of optical colors to the quench rate
and found that for example at a fixed (transitional) g-i color there is very little
sensitivity to the quench rate in the \dn vs. \hda plane. In other words
the optical continuum given by smooth exponential models 
and that produced by our quenched models do not yield significantly
different values of \hda and \dn since all three are tracing similar
timescales. This means that A$_z$ derived from the indices and
the optical continuum should be reasonably accurate even for quenched models.
A self-consistent approach is certainly warrented in a future study.
A small fraction of the $A_z$ extinctions are negative--for these
we assumed zero extinction.

As a check we repeated the calculation
using extinctions obtained by \cite{brinchmann04} from the Balmer decrement (and
additional lines), again with a \cite{calzetti94} extinction law. 
A significant fraction of the galaxies in the transitional color
bins do not have lines strong enough to support this technique. For those we continued
to use the continuum based extinctions. The results of Method 3 are virtually
identical as are the
values of the quenching rates. This is comforting, since the \cite{calzetti94}
law assumes a particular relationship between H$\alpha$ and UV
extinction which was derived for starburst galaxies.

Removal of more extincted, presumably star forming galaxies would likely result in
lower values of $\phi(M_r, NUV-r)$ and $dy/dt$, yielding even lower values of the
transitional mass flux. 

\subsubsection{Color Variation}

We chose a transitional color bin of $4.0<NUV-r<4.5$. Is our result
dependent on this choice? If our approach is sensible, continuity requires
that the mass flux
calculated from neighboring color bins should be similar,  as long as 
we are probing truly transitional objects. Differences would arise for
two reasons. First, redder transitional colors probe slightly older 
timescales. If there has been any evolution in the mass flux,
then this would produce a trend. It is likely that the mass flux was larger in the past (preliminary
analysis using AEGIS data \cite{martin06c} indicates some evolution).
Secondly, objects that reach the
red sequence could experience episodic residual star formation
events \cite{yi05} that move them back into the transition region or the bluest
part of the red sequence, increasing the apparent mass flux.

We have repeated the analysis, using Method 3, on two neighboring color bins.
The results for $NUV-r=$3.75, 4.25[baseline], and 4.75 are, respectively
$\dot{\rho}_T=0.026, 0.033, 0.038$ M$_\odot$ yr$^{-1}$ Mpc$^{-3}$. 
The trend is consistent with both of the effects discussed above.
However, the trend is modest, suggesting that to first
order the basic method is sound.

\subsection{Transition Region Galaxies, AGN Feedback, and Mergers}

\begin{figure}
\plotone{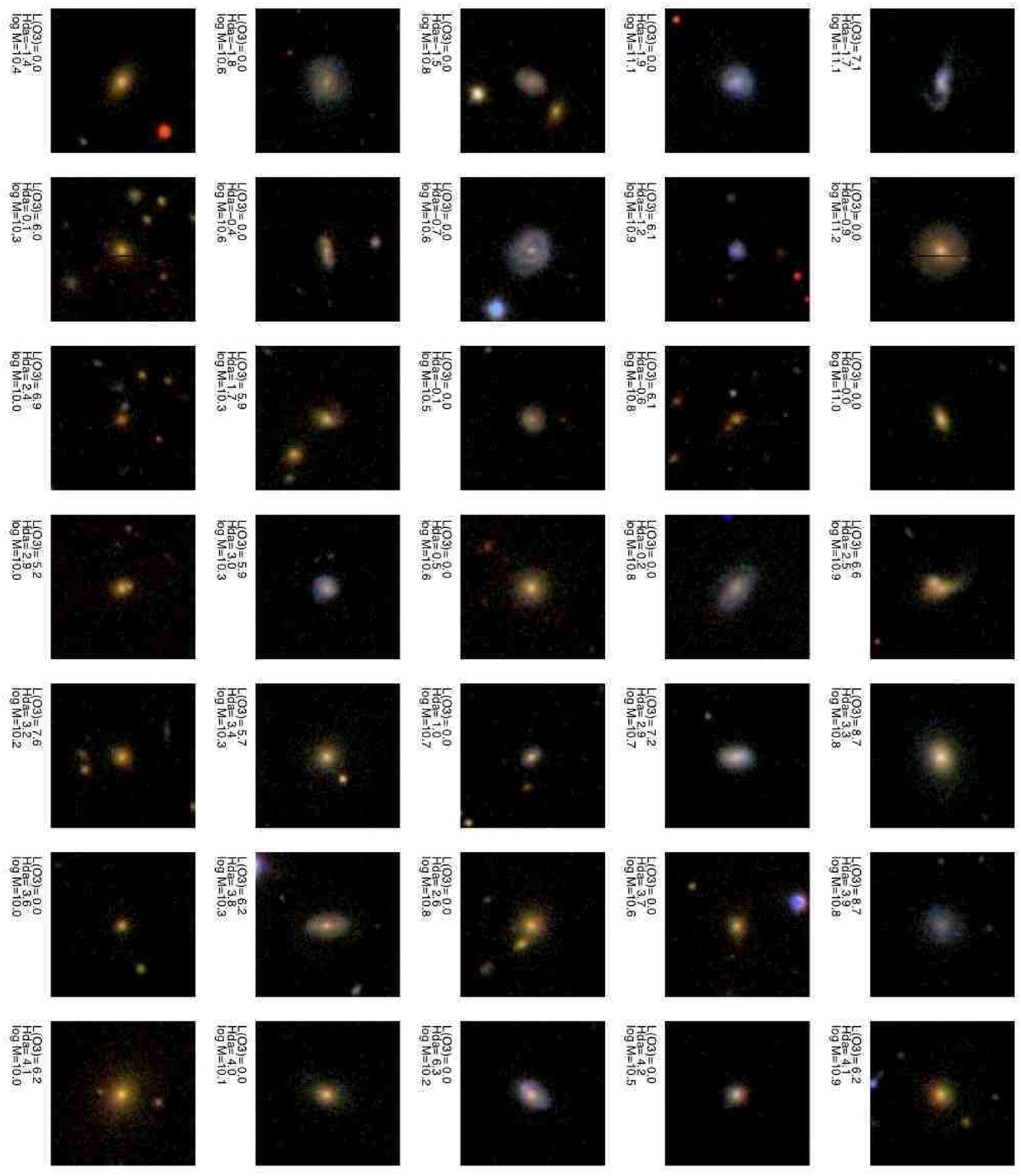}
\caption{Sample atlas of SDSS 45 arcsecond square thumbnails of transition galaxies.
Each row is a fixed M$_r$ bin, brightest to dimmest going top to bottom,
Images are ordered left to right by \hda (a proxy in our model for quench rate),
with \hda (and quench rate) increasing to the right. Log of stellar mass
and \hda are given.
\label{fig_atlas}}
\end{figure}

What kind of galaxies occupy the transitional color region of the \hess?
We show in Figure \ref{fig_atlas} a small atlas of transitional galaxy
thumbnails from the SDSS. The galaxies are organized by mass and
\hda. 

One interesting and suggestive property is the preponderence of AGNs.
We have used the value added dataset of AGN properties generated by
\cite{heckman04} to explore this.

\begin{figure}
\plottwo{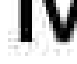}{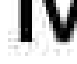}
\caption{\hess~with AGN fraction plotted as greyscale in each bin. Fraction of galaxies
in the bin that have AGN or AGN-starburst composite lines based on Heckman et al. (2004). LEFT:
No extinction correction. RIGHT: Extinction corrected.
Contours are spaced logarithmically in 10 even steps from $\phi=10^{-5}-10^{-3}$ (0.2 dex per step).
\label{fig_agnfrac_hess}}
\end{figure}

\begin{figure}
\plottwo{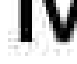}{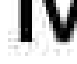}
\caption{\hess~with AGN [OIIII] luminosity plotted as color map in each bin based on Heckman et al. (2004). LEFT:
No extinction correction. RIGHT: Extinction corrected.
Contours are spaced logarithmically in 10 even steps from $\phi=10^{-5}-10^{-3}$ (0.2 dex per step).
Numbers in each bin are number of galaxies.
\label{fig_lumo3agn_hess}}
\end{figure}

We consider two ensemble properties. \cite{heckman04} calculated the luminosity in the
OIII] line for galaxies that showed line ratios consistent with AGN or composite AGN/star formation
ionizations. In Figure \ref{fig_agnfrac_hess} we show the number fraction of galaxies
with some AGN emission in each color-magnitude bin of the \hess. In Figure \ref{fig_lumo3agn_hess}
we show the mean value of $L(OIII])$ for the AGNs in each bin. In each of these cases,
although it does not make a signficant qualitative difference, we have used the
extinction-corrected \hess. We assume that AGN continuum in these Seyfert II objects
is neglible even in the NUV. \footnote{If this is not the case, some
galaxies at transitional colors are red sequence ``interlopers''. In principal
the mass flux could be even lower when these objects are removed, although the correction
could also add some galaxies from bluer bins.}

Figure \ref{fig_agnfrac_hess} is particularly striking, showing that the AGN fraction rises
precisely in the transitional color region between the blue and red sequence. 
This is of course extremely suggestive of an important role for AGN feedback
\citep{springel05} in quenching star formation more rapidly than
simple gas exhaustion. 

\begin{figure}
\plottwo{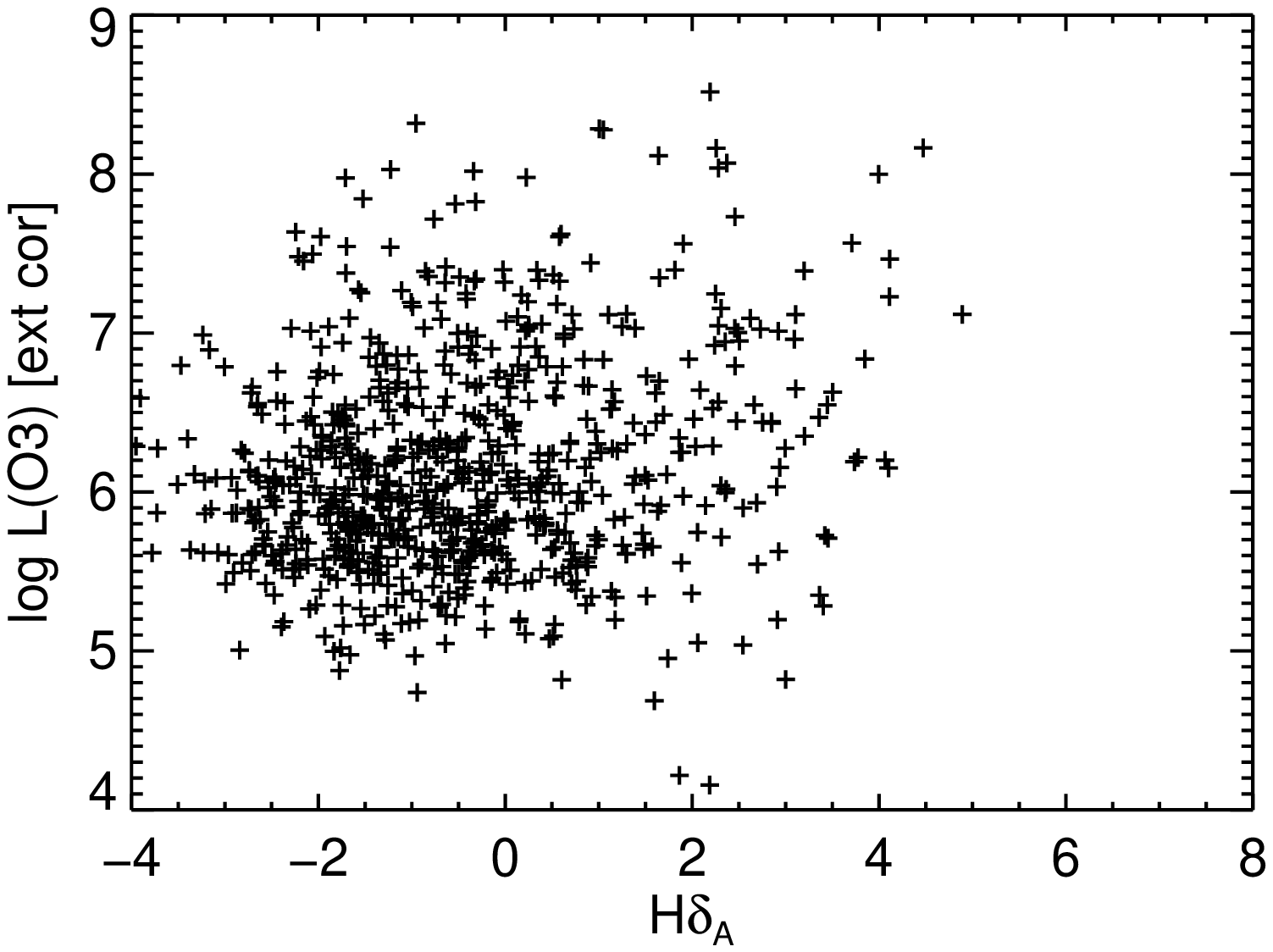}{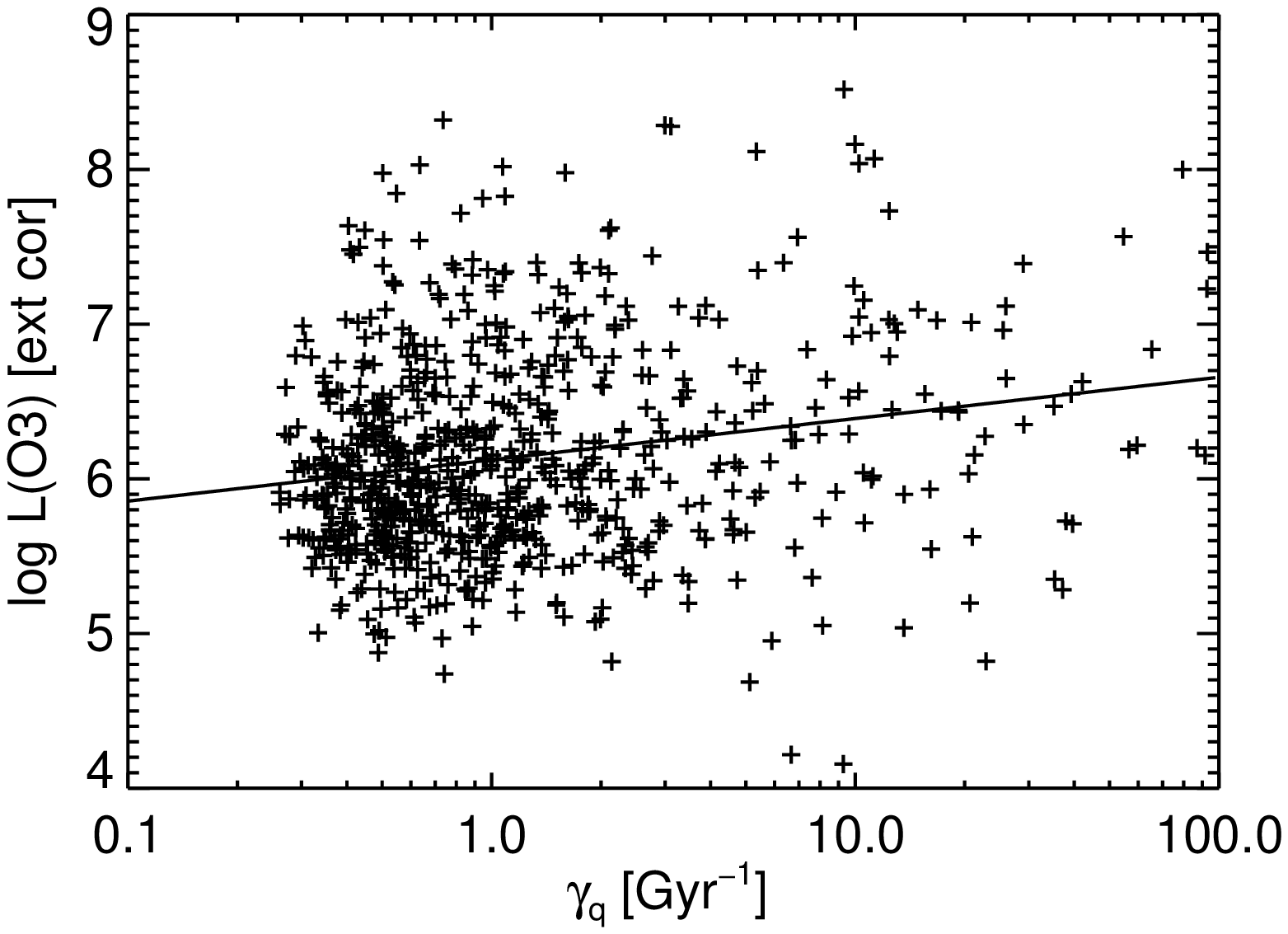}
\caption{We show for each galaxy in the range $-23<M_r<-20$
$L(OIII])$ vs. \hda [LEFT] and the derived quench rate $\gamma$ [RIGHT].
Color range is 3.75$<$NUV-r$<$4.75.
\label{fig_agn_gam}}
\end{figure}

\begin{figure}
\plotone{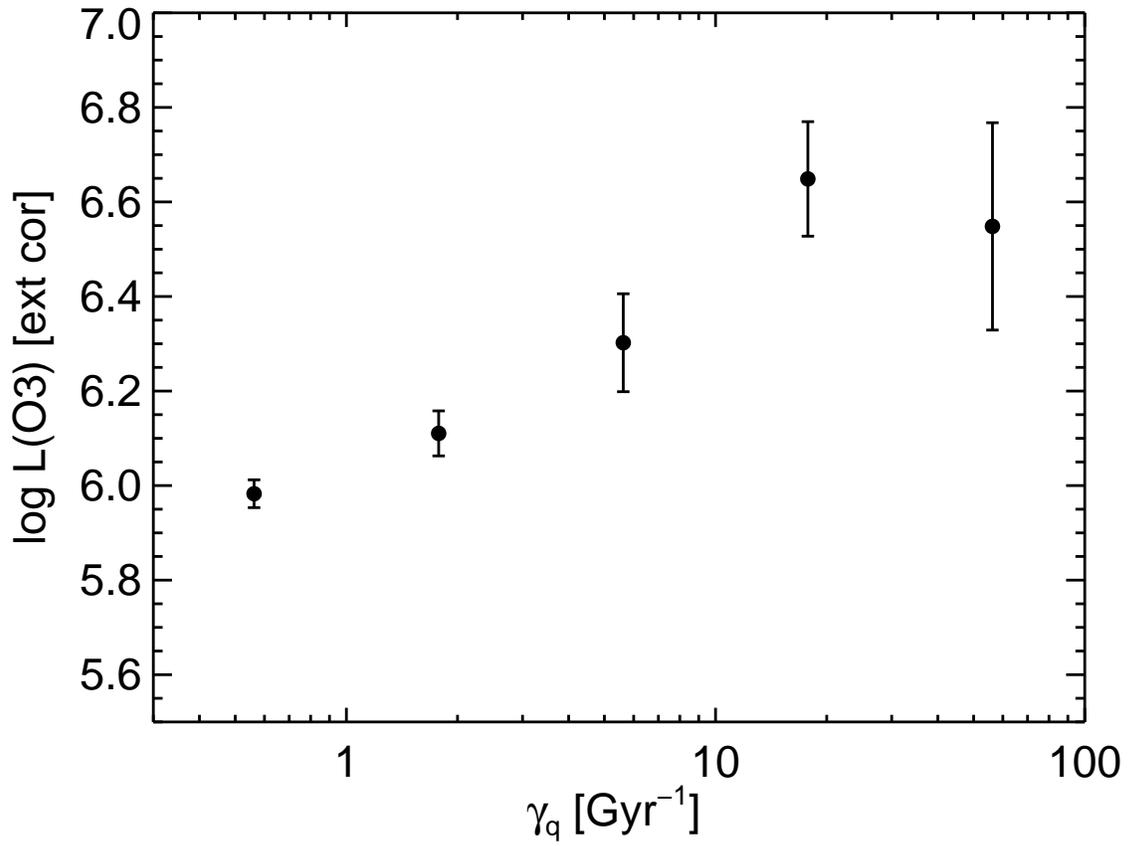}
\caption{Median of $\log L_{[OIII]}$ vs. $\gamma$ in bins of $\Delta\gamma=0.5$.
Color range is 3.75$<$NUV-r$<$4.75. Absolute magnitude range is $-23<M_r<-20$.
\label{fig_agn_gam_bin}}
\end{figure}

In the context of this work, we can ask whether the derived model parameters,
in particular the quench rate, is correlated with AGN properties, such as luminosity.
In Figure \ref{fig_agn_gam}, we show for each galaxy in the range $-23<M_r<-20$
$L(OIII])$ vs. \hda and the derived quench rate $\gamma$. There is considerable scatter.
Linear regression gives for $\log{L(OIII])}=a \log{\gamma} + b$, $a=0.37\pm 0.05$, with 
a correlation coefficient of only 0.32.  
The trend is most evident in Figure \ref{fig_agn_gam_bin}, where the 
median luminosity is calculated in five $\gamma$ bins. 

There is a modest trend of $L(OIII])$ increasing with $\gamma$.
Luminosity nominally traces accretion rate, which is likely to be higher in
the gassier galaxies nearer the blue sequence. Higher \hda galaxies have more recently
lost their gas fueling star formation, and there may be a time delay for that
to starve the central engine.  Thus the strongest statement that
can be made is that galaxies with AGNs of higher luminosity tend to
show higher quench rates. 

\begin{figure}
\plotone{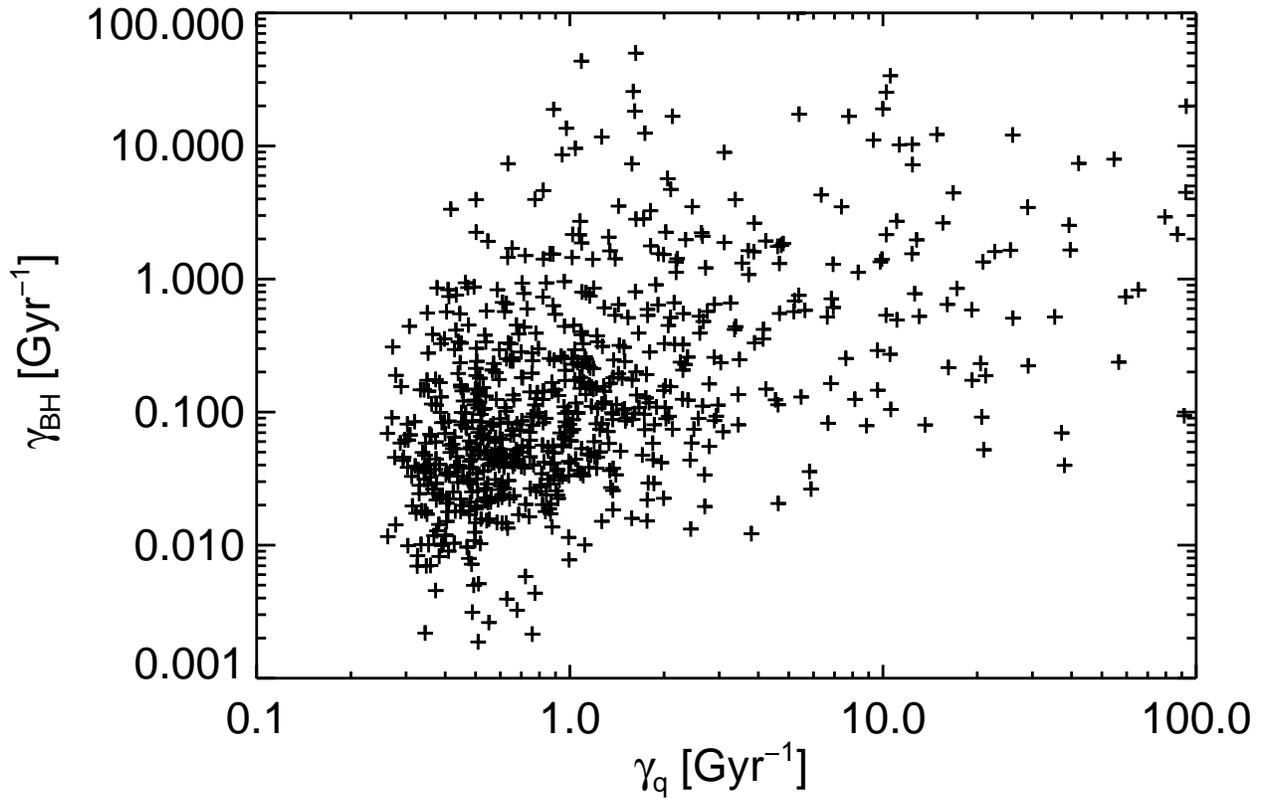}
\caption{Black hole growth rate $\gamma_{BH}$ (see text) vs. $\gamma_{quench}$.
Color range is 3.75$<$NUV-r$<$4.75. Absolute magnitude range is $-23<M_r<-20$.
\label{fig_gambh}}
\end{figure}

\cite{heckman04} computed a specific growth rate for central black holes
by using $L(OIII])$ to obtain a rough bolometric luminosity and
the central velocity dispersion and the $M-\sigma$ relation \citep{tremaine02}
to obtain the black hole mass. The result is

\begin{equation}
{{\dot{M}_{BH}} \over {M_{BH}}} = \gamma_{BH} = 10^{-7.8} L(OIII]) {\sigma_{200}}^4 {\rm Gyr^{-1}}
\end{equation}

which we compared with the derived quench rate in Figure \ref{fig_gambh}.
There is a clear trend of increasing $\gamma_{BH}$ with increasing $\gamma$.
But there is considerable dispersion, and the current black hole growth
rate is typically much less than the quench rate. Most of the AGNs in the transition
region are accreting far below the Eddington limit \citep{heckman04}. This suggests that if the
processes of black hole growth and star formation are linked, the timescale
for the quenching of black hole growth is considerably shorter than
the quenching of star formation. 

\begin{figure}
\plotone{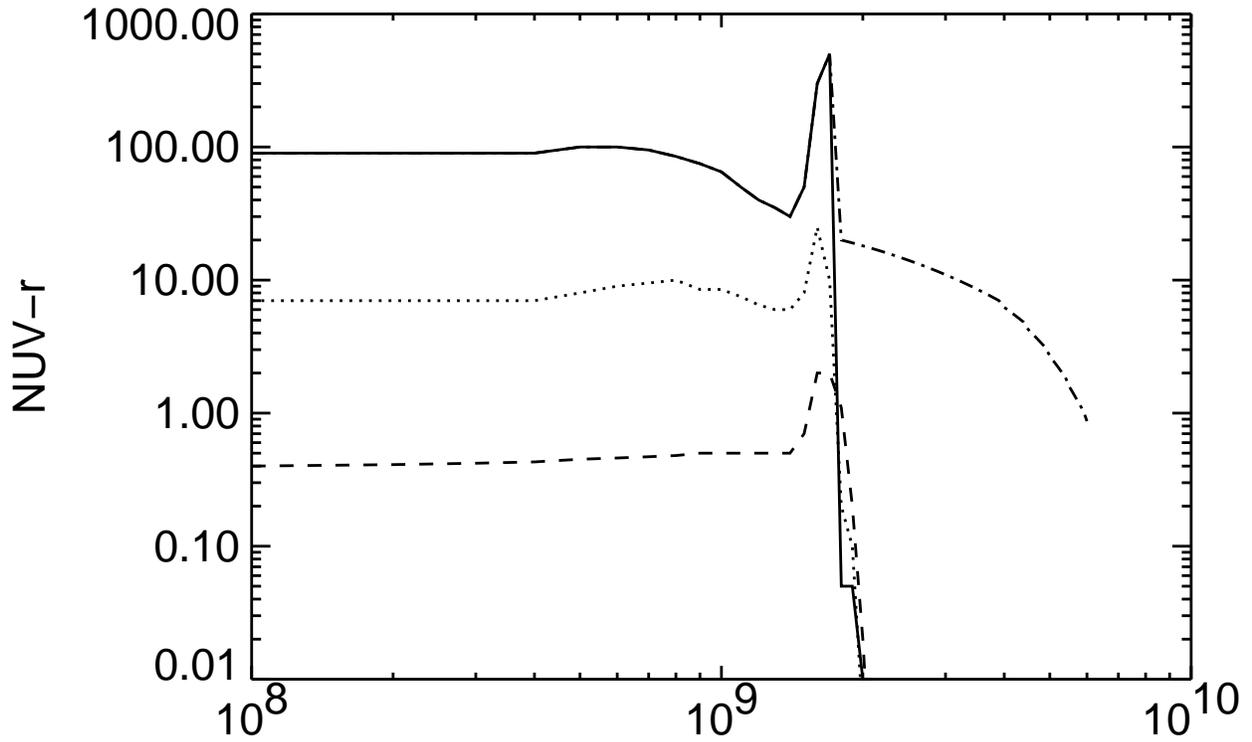}
\caption{Star formation history of simulated major mergers with and without AGN feedback from DiMatteo et al. (2005). 
SOLID: v(rotation)=320 km/s, with feedback. 
DOT-DASHED: v(rotation)=320 km/s, no feedback. 
DOTTED: v(rotation)=160 km/s, with feedback.
DASHED: v(rotation)=80 km/s, with feedback.
\label{fig_sfr_feedback}}
\end{figure}

\begin{figure}
\plotone{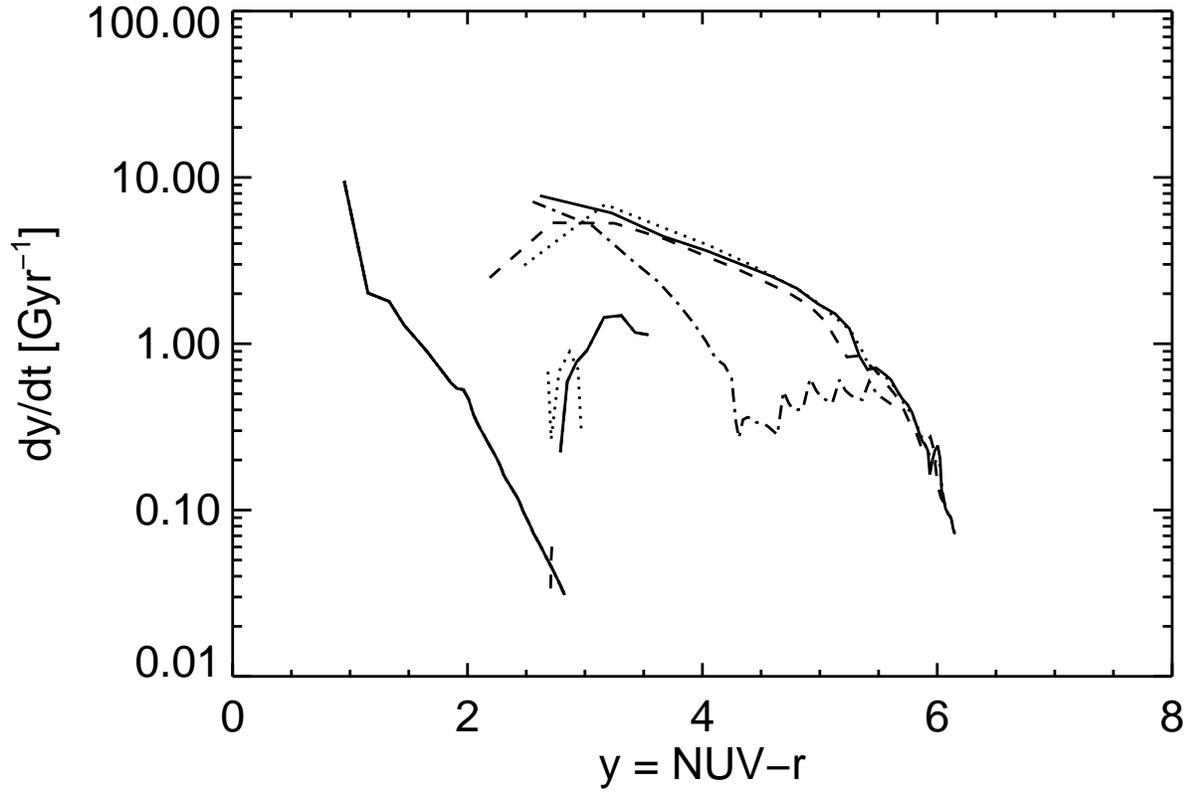}
\caption{Color derivative evolution for the 4 DiMatteo et al. (2005) models.
SOLID: v(rotation)=320 km/s, with feedback. 
DOT-DASHED: v(rotation)=320 km/s, no feedback. 
DOTTED: v(rotation)=160 km/s, with feedback.
DASHED: v(rotation)=80 km/s, with feedback.
\label{fig_feedback_dydt}}
\end{figure}

\begin{figure}
\plotone{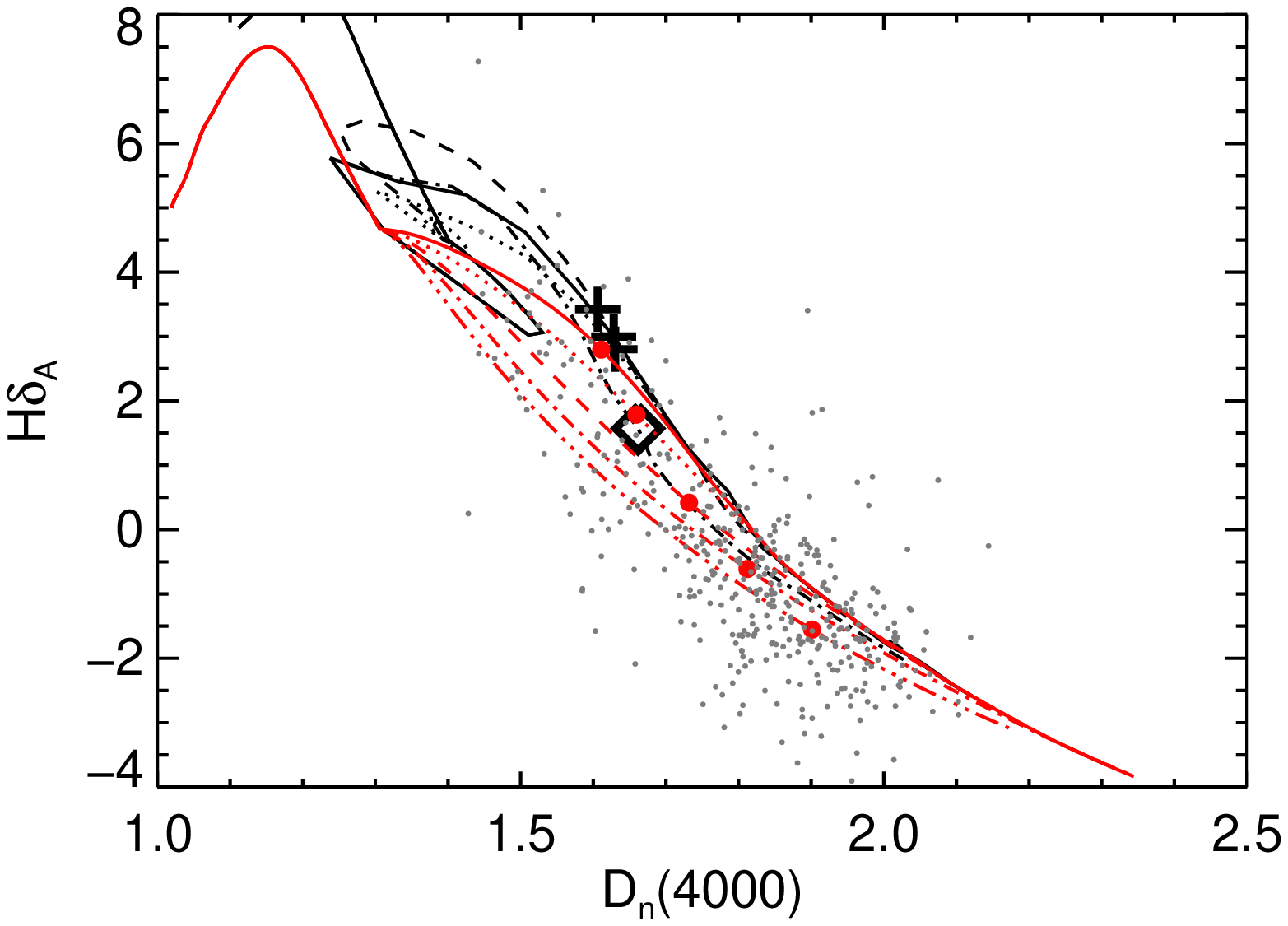}
\caption{Feedback models evolution in \hda and \dn plane (black lines).
SOLID: v(rotation)=320 km/s, with feedback. 
DOT-DASHED: v(rotation)=320 km/s, no feedback. 
DOTTED: v(rotation)=160 km/s, with feedback.
DASHED: v(rotation)=80 km/s, with feedback.
Large symbols show values at transition color NUV-r=4.25. Plusses
show models with feedback, diamond shows model 2 with no feedback.
Red curves show simple quenched models for reference.
\label{fig_feedback_hda_d4000}}
\end{figure}

\cite{springel05} modeled the effects of AGN feedback on the star formation
history of a post-merger galaxy. Using their star formation rate histories,
we have modelled the color and index evolution in the same fashion as our
simple quenched models. In order to make a consistent comparison, we have
assumed that their star formation history follows 10 Gyr of constant star formation
rate at the starting rate. Figure \ref{fig_sfr_feedback} shows
the star formation history. Figure \ref{fig_feedback_dydt} shows the color
derivative $dy/dt$ for the four models (3 with AGN feedback and various \ref{fig_feedback_hda_d4000} shows the
loci of models on the \hda-\dn plane, compared with the quenched models and
the data as in Figure \ref{fig_hda_d4000_data}. The crosses and diamond (for the no feedback model) indicate the points when
$NUV-r=4.25$. Three of the models have black hole feedback (and three values of
dark halo mass), while the fourth shows the most massive object without
the feedback effects. Interestingly, there is reasonable overlap between
the numerical models and the simple quenched model. These models exhibit
a slightly higher value of \hda.
A partial reason is that the numerical models have a large starburst
just prior to quenching, which tends to elevate \hda (and lower \dn) at a given
NUV-r color for the fading population. The color derivative for these
models at the transitional color is $dy/dt\simeq 3$ Gyr$^{-1}$ with feedback
and $dy/dt\simeq 0.6$ Gyr$^{-1}$ without. A moderate fraction of galaxies
show the combination of color, \hda and \dn predicted by the merger/feedback models,
while many more show the older indices predicted for mergers and no feedback. 
But a sizeable fraction show even older indices inconsistent with the numerical histories.
Using these tracks to interpret our data leads to lower values of $dy/dt$ for a given
combination of index at our transitional color, and a correspondingly lower
mass flux than we derived above. 

\begin{figure}
\plotone{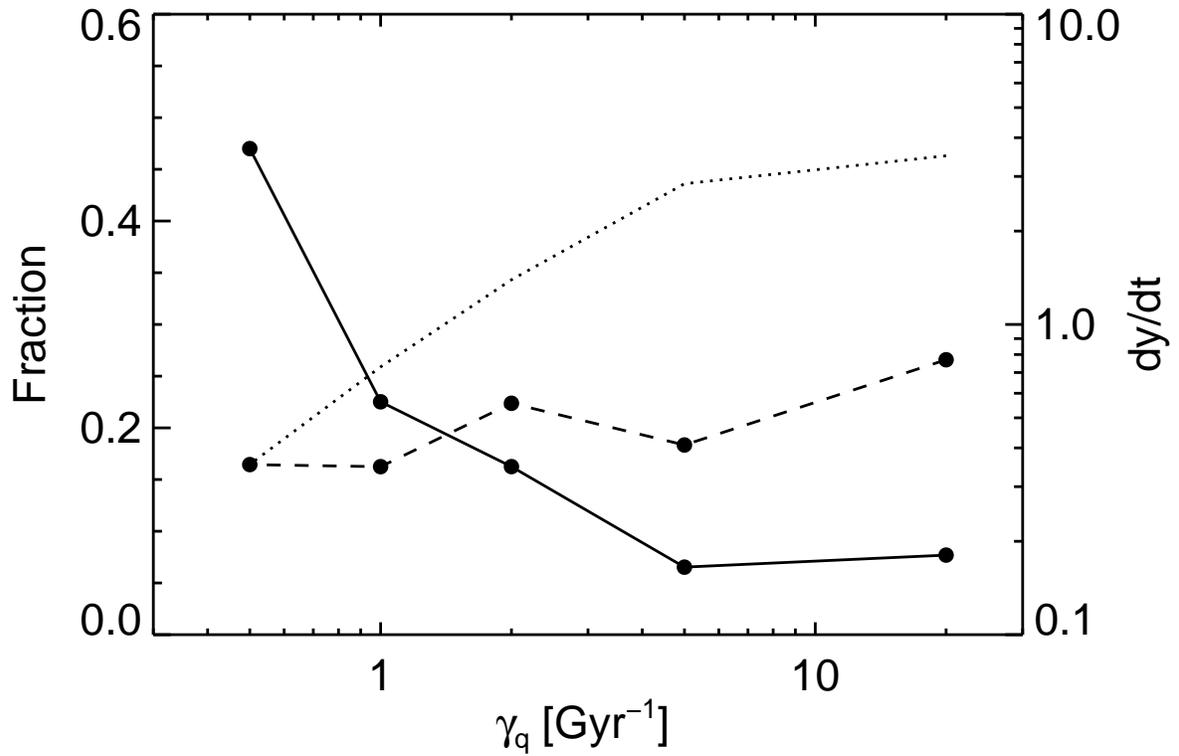}
\caption{Volume corrected fraction of galaxies in 5 $\gamma$ bins shown in Figure \ref{fig_hda_d4000_bin}.
Galaxies have $-23<M_r<-20$. Solid line shows unweighted fraction in each bin.
Dashed line shows fraction weighted by dy/dt, which corrects for the
shorter residence time of fast-quenching galaxies. Dotted line shows
the color derivative dy/dt in each $\gamma$ bin.
\label{fig_method3_bin}}
\end{figure}

Roughly speaking, about 8\% of galaxies (bin one of Figure \ref{fig_method3_bin}) 
show properties consistent with the
rapid quenching in these major merger models with AGN feedback. Some
10-20\% show properties like the mergers without feedback, and the balance
show even slower quenching rates. The majority of galaxies do not exhibit the
rapid quench rates predicted by strong AGN feedback. However, it is important to
remember that rapid quenches transition rapidly. From Figure \ref{fig_method3_bin}
(last column), we see that when we correct for the residence time in the transition
zone, 50\% of galaxies have quench rates $\gamma \ge 5$ Gyr$^{-1}$. 

Finally, we can compare these statistics to measurements of the cosmic
merger rate. For the redshift range $0.4<z<0.8$, \cite{conselice03} obtain
a stellar mass accretion rate due to major mergers ($M_B<-19$) of 
$1.8\times 10^{-3}$ M$_\odot$ Mpc$^{-3}$ Gyr$^{-1}$, which is about
7\% of the total mass flux we measure.  

\section{Summary}

We have introduced a new quantity, the mass flux density of galaxies evolving
from the blue sequence to the red sequence. We developed a simple technique
for constraining this mass flux that exploits the excellent separation of red and blue sequences in the NUV-r band \hess. 
We used the volume corrected number density
in the extinction-corrected UV-optical color-magnitude distribution, the stellar age indexes \hda and \dn,
and a simple prescription for spectral evolution using a quenched star formation history.
Our estimated mass flux, $\dot{\rho_T}=0.033$ M$_\odot$ yr$^{-1}$ Mpc$^{-3}$, although strictly
an upper limit, compares favorably with
estimates of the average mass flux that we make based on the optical
and UV data. Galaxies in the transition zone are preferentially AGNs, although 
we find at best weak evidence that the quench rates are higher in higher luminosity AGNs.

We note that our technique could be applied with alternative evolutionary clocks,
such as morphology. Our approach using a single color and spectral index was
simple, but a more refined technique would use the entire spectral energy distribution
in addition to any relevant spectral indices. Our simple approach is fairly
easy to apply to deep galaxy surveys, future work will study the evolution of the mass
flux as well as its dependence on galaxy density and environment.





\acknowledgments

GALEX (Galaxy Evolution Explorer) is a NASA Small Explorer, launched in April 2003.
We gratefully acknowledge NASA's support for construction, operation,
and science analysis for the GALEX mission,
developed in cooperation with the Centre National d'Etudes Spatiales
of France and the Korean Ministry of 
Science and Technology. We also thank the referee for excellent comments.



{\it Facilities:} \facility{GALEX}, \facility{SDSS}

\clearpage
\begin{deluxetable}{rrrrrrrrrr}
\tabletypesize{\scriptsize}
\tablecaption{Mass Flux Table Method 1 \label{tab_method1}}
\tablewidth{0pt}
\tablehead{
\colhead{M$_r$} & 
\colhead{$\log{M_*}$} & 
\colhead{$\phi$} & 
\colhead{$\#$} &
\colhead{$\overline{H\delta_A}$} & 
\colhead{$\overline{D_n(4000)}$} &
\colhead{$\gamma$} & 
\colhead{dy/dt} & 
\colhead{$\dot{\phi}$} &
\colhead{$\dot{\rho}_B$}
}
\startdata
   -23.25&     0.00&  3.30e-08&    1&     -1.13&      1.73&      1.20&      0.93&  1.23e-07&    0.0000\\
   -22.75&    11.38&  1.39e-06&   42&     -0.24&      1.78&      1.25&      0.96&  5.34e-06&    0.0006\\
   -22.25&    11.23&  8.80e-06&  172&      0.27&      1.75&      1.73&      1.29&  4.56e-05&    0.0039\\
   -21.75&    11.07&  3.22e-05&  358&     -0.02&      1.75&      1.57&      1.19&  1.53e-04&    0.0089\\
   -21.25&    10.86&  6.79e-05&  403&      0.41&      1.71&      2.21&      1.58&  4.29e-04&    0.0156\\
   -20.75&    10.65&  9.53e-05&  304&      0.63&      1.70&      2.58&      1.77&  6.76e-04&    0.0150\\
   -20.25&    10.47&  1.25e-04&  215&      1.04&      1.66&      3.81&      2.28&  1.14e-03&    0.0170\\
   -19.75&    10.26&  1.33e-04&  114&      1.40&      1.60&      6.54&      2.94&  1.57e-03&    0.0143\\
   -19.25&    10.04&  1.13e-04&   50&      1.47&      1.57&      8.15&      3.17&  1.43e-03&    0.0078\\
   -18.75&     9.77&  1.64e-04&   34&      2.33&      1.50&     18.22&      3.58&  2.35e-03&    0.0069\\
   -18.25&     9.42&  1.01e-04&   10&      1.70&      1.53&     11.80&      3.45&  1.40e-03&    0.0018\\
   -17.75&     9.37&  1.88e-04&    6&      2.27&      1.43&     20.06&      3.58&  2.69e-03&    0.0031\\
\enddata
\end{deluxetable}


\clearpage

\begin{deluxetable}{rrrrrrrrrr}
\tabletypesize{\scriptsize}
\tablecaption{Mass Flux Table Method 2 \label{tab_method2}}
\tablewidth{0pt}
\tablehead{
\colhead{M$_r$} & \colhead{$\log{M_*}$} & \colhead{$\phi$} &\colhead{$\#$} & \colhead{$\overline{H\delta_A}$} & \colhead{$\overline{D_n(4000)}$} &
\colhead{$\gamma$} & \colhead{dy/dt} & \colhead{$\dot{\phi}$} & \colhead{$\dot{\rho}_B$}
}
\startdata
   -23.75&    11.63&  1.84e-07&    5&     -0.61&      1.81&      0.95&      0.73&  5.38e-07&    0.0001\\
   -23.25&    11.49&  1.08e-06&   22&     -1.05&      1.85&      0.72&      0.53&  2.30e-06&    0.0004\\
   -22.75&    11.32&  6.53e-06&  100&     -0.69&      1.84&      0.84&      0.63&  1.65e-05&    0.0017\\
   -22.25&    11.12&  2.20e-05&  223&     -0.72&      1.83&      0.87&      0.66&  5.83e-05&    0.0038\\
   -21.75&    10.93&  4.02e-05&  259&     -0.66&      1.82&      0.91&      0.69&  1.11e-04&    0.0047\\
   -21.25&    10.73&  5.63e-05&  224&     -0.67&      1.82&      0.91&      0.69&  1.57e-04&    0.0042\\
   -20.75&    10.50&  7.42e-05&  172&     -0.52&      1.81&      0.97&      0.75&  2.22e-04&    0.0035\\
   -20.25&    10.25&  6.80e-05&   81&     -0.18&      1.76&      1.41&      1.08&  2.94e-04&    0.0026\\
   -19.75&    10.00&  6.94e-05&   41&      0.25&      1.71&      2.13&      1.53&  4.25e-04&    0.0021\\
   -19.25&     9.85&  6.06e-05&   22&     -0.19&      1.73&      1.64&      1.24&  3.01e-04&    0.0011\\
   -18.75&     9.52&  7.22e-05&   10&      1.69&      1.56&      9.86&      3.33&  9.61e-04&    0.0016\\
   -18.25&     9.32&  1.16e-04&    7&      0.93&      1.60&      5.08&      2.64&  1.23e-03&    0.0013\\
   -17.75&     9.01&  1.18e-04&    2&      3.01&      1.44&     25.94&      3.51&  1.66e-03&    0.0009\\
\enddata
\end{deluxetable}


\clearpage


\begin{deluxetable}{rrrrrrrrrrr}
\tabletypesize{\scriptsize}
\tablecaption{Mass Flux Table Method 3 \label{tab_method3}}
\tablewidth{0pt}
\tablehead{
\colhead{M$_r$} & \colhead{$\log{M_*}$} & \colhead{$\phi$}&\colhead{$\#$}  & \colhead{$\overline{H\delta_A}$} & \colhead{$\overline{D_n(4000)}$} &
\colhead{$\overline{\gamma}$} & \colhead{$\overline{dy/dt}$} & \colhead{$\dot{\phi}$} & \colhead{$\dot{\rho}_B$} & \colhead{$\sigma[\dot{\rho}_B]$}
}
\startdata
 -23.75&    11.63&  1.84e-07&    5&     -0.61&      1.81&      1.00&      0.82&  6.08e-07&    0.0001&    0.0001\\
   -23.25&    11.49&  1.08e-06&   22&     -1.05&      1.85&      0.85&      0.70&  3.03e-06&    0.0005&    0.0001\\
   -22.75&    11.32&  6.53e-06&  100&     -0.69&      1.84&      1.02&      0.85&  2.22e-05&    0.0023&    0.0003\\
   -22.25&    11.12&  2.20e-05&  217&     -0.72&      1.83&      1.05&      0.87&  7.71e-05&    0.0051&    0.0005\\
   -21.75&    10.93&  4.02e-05&  252&     -0.66&      1.82&      1.07&      0.94&  1.51e-04&    0.0064&    0.0006\\
   -21.25&    10.73&  5.63e-05&  211&     -0.67&      1.82&      1.02&      0.87&  1.97e-04&    0.0053&    0.0005\\
   -20.75&    10.50&  7.42e-05&  154&     -0.52&      1.81&      1.19&      1.05&  3.11e-04&    0.0050&    0.0004\\
   -20.25&    10.25&  6.80e-05&   77&     -0.18&      1.76&      1.56&      1.37&  3.74e-04&    0.0033&    0.0004\\
   -19.75&    10.00&  6.94e-05&   40&      0.25&      1.71&      1.97&      1.69&  4.70e-04&    0.0024&    0.0004\\
   -19.25&     9.85&  6.06e-05&   21&     -0.19&      1.73&      1.37&      1.50&  3.64e-04&    0.0013&    0.0003\\
   -18.75&     9.52&  7.22e-05&    9&      1.69&      1.56&      3.68&      2.90&  8.37e-04&    0.0014&    0.0005\\
\enddata
\end{deluxetable}

\clearpage



\clearpage

\clearpage

\clearpage

\clearpage

\clearpage

\clearpage

\clearpage

\clearpage

\clearpage

\clearpage

\clearpage

\clearpage

\clearpage

\clearpage

\clearpage

\clearpage

\clearpage

\clearpage

\clearpage

\clearpage

\clearpage

\clearpage

\end{document}
